\documentclass[singlecolumn,a4paper,aps,amsfont,preprintnumbers,balancelastpage,10pt]{revtex4-2}
\usepackage{amsmath}
\usepackage{amssymb}
\usepackage{amsfonts}
\usepackage{fontenc}
\usepackage[mathcal]{eucal}
\usepackage{bm}
\usepackage{lmodern}
\usepackage{epstopdf}
\usepackage{epsfig}
\usepackage{multirow}
\usepackage{dcolumn}
\usepackage{array}
\usepackage{rotating}
\usepackage[breaklinks]{hyperref}
\usepackage{makecell}
\usepackage{graphicx}
\usepackage{subfigure}
\DeclareMathAlphabet{\mathpzc}{OT1}{pzc}{m}{it}
\begin{document}
	\title{Tidal deformability and compactness of neutron stars and massive \\ pulsars from semi-microscopic equations of state}
	\author{W. M. Seif$^{1,2,3,(*)}$ and A. S. Hashem$^{1}$}
	\affiliation{\footnotesize $^{1}$Cairo University, Faculty of Science, Department of Physics, 12613 Giza, Egypt\\
	$^{2}$Joint Institute for Nuclear Research, 141980 Dubna, Russia\\$^{3}$The Academy of Scientific Research and Technology, 4262104 Cairo, Egypt\\
	$^{(*)}$\href{mailto:wseif@sci.cu.edu.eg}{{wseif@sci.cu.edu.eg}}}

\begin{abstract}
Tidal deformability measures how NS can comfortably deform as a response to an applied tidal field. We use updated constraints on the mass, radius, and tidal deformability of neutron star (NS) objects and pulsars to examine nuclear equations of state (EOS) based on realistic finite-range M3Y nucleon-nucleon interaction, which have been successfully used to describe low- and high-dense nuclear matter (NM). We then employ these EOSs to examine the impact of tidal deformability and compactness of NSs on their structure. We found that the EOSs from CDM3Y-230 to CDM3Y-330 characterized with the saturation incompressibility ${K}_{0}=230-330$ MeV together yield more limited ranges of tidal deformability and radius for NS objects than their experimentally inferred ranges. For light NS ($M<\text{M}_{\odot}$), both ${k}_{2}$ and $\mathpzc{C}$ decreases upon decreasing the NS mass, which enhances its tidal deformability. The stiffness of the NS core matter has shown a minor effect on the tidal deformability of such NS ($M<\text{M}_{\odot}$). An opposite behavior is obtained as an increase in the tidal Love number but a decrease in the more effective compactness of NS ($M>\text{M}_{\odot}$), upon increasing (decreasing) the stiffness of the employed EOS (its mass). This appears as enhanced tidal deformability indicated at a larger radius for NS of stiffer NM and for the lighter NS above $\text{M}_{\odot}$. Unified description of some correlations between tidal deformability, tidal Love number, and NS compactness is provided independent of the details of the considered EOS.  

\end{abstract}	
	
\maketitle

\section{\label{sec:level1}Introduction}

 Observations of neutron stars have shown that they are extremely dense structures, in which all four fundamental forces play a combined role and its core matter can reach densities up to many times the nuclear saturation density \cite{HaenselPotekhin2007,TanKhoa2021}. Neutron stars are proposed to be the most compact astronomical objects next to black holes. As they are composed of baryonic and leptonic matter under extreme physical conditions, they provide rich potential research area in modern nuclear astrophysics. The coalescence of binary massive compact neutron star (NS) systems acts as a strong astrophysical source of gravitational waves (GW) that can be detected by advanced ground-based detectors \cite{TaylorWeisberg1982,CutterApostolatos1993,MalikAlam2018}, which are sensitive to indicate tidal deformability. Most of the energy of binary black hole and binary NS merging due to their collision at high density is expected to appear in the form of GW. The observed GW signals from NS merger are expected to be sensitive to the behavior of dense NS matter and its equation of state (EOS) \cite{Faber2009,Duez2010}. Consequently, it can be considered as a valuable tool to assess the behavior of asymmetric nuclear matter (ANM) and its related quantities at the extreme conditions that cannot be implemented in laboratory experiments or naturally on earth. For instance, the extracted restrictions on tidal deformability, or polarizability, from observed GW starting from the GW170817 binary NS merger \cite{Abbottetal2018} have been used to constrain the stellar radii, the mass-radius relation, and stellar compactness, in addition to link them to the nuclear EOS \cite{Piekarewicz2019,AnnalaGorda2018,Fattoyev2018}. Also, the tidal Love number that is related to the tidal deformability of NS correlates with the crustal equation of state \cite{Binnington2009,DamourNagar2012}. The tidal deformability characterizes the extent of deformation of NS caused by the associated tidal field \cite{HindererLackey2010}. During the late inspiral of a coalescing binary NS system, tidal deformation causes each NS to possess a non-zero quadrupole moment under the mutual gravitational field created by the merging NS partners \cite{HindererLackey2010,Hinderer2008}. After a few indirect evidences and next to the first direct observation of GW from a binary black hole coalescence in 2016 by the LIGO and Virgo Collaborations \cite{Abbottetal2016}, the direct detection of GW from binary NS mergers has been started by the GW170817 signal in 2017, which identified by LIGO and VIRGO laser interferometers \cite{Abbottetal2017}, and followed by a short burst of $\gamma$ rays detected by the Fermi Gamma-Ray Burst Monitor (GBM) \cite{Goldsteinetal2017} and by the INTErnational Gamma-Ray Astrophysics Laboratory (INTEGRAL) \cite{Savchenkoetal2017}. This launched a new era of astrophysics, multi-messenger astronomy, and fundamental physics. In addition to GW and $\gamma$ ray bursts, the subsequent radiation related to the binary black-hole and NS mergers includes observed X-ray and infrared radiations, visible light and other electromagnetic bands, in addition to expected high-energy neutrinos, which offers a rich database to generate new findings. \par 
Next to the GW170817 event, a massive in-depth investigation of the EOS and its related quantities had been focused on the NS systems and gravitational waves. One of the main ambitious challenges for such studies is to reach reliable EOS for the different regions of the core and crust of NS starting from the principles of nuclear theory, which can simultaneously describe the NS matter and its structure in consistence with the theories of plasma in the core and surface NS layers as well as the atomic structure influences in crust. Also, the different nuclear forces, the many-body interactions at high density and temperature, and the quark matter, in additions to the elementary particle production and interactions are expected to show roles. Towards this, many hadronic \cite{HindererLackey2010,PostnikovPrakash2010,MoustakidisGaitanos2017,KumarBiswal2017,SeifHashem2024} EOSs have been examined. Also, approximate polytropic and simple phenomenological EOSs had been used to describe the individual NS layers \cite{Binnington2009,DamourNagar2009}. For instance, the sensitivity of the tidal deformability and Love number to the considered EOS has been systematically studied \cite{FlanaganHinderer2008,Binnington2009,HindererLackey2010,DamourNagar2009}, either analytically or in numerical simulations, to place upper and lower limits on both the tidal deformability and EOS related quantities, as well as to investigate the gravitational waveform of a binary compact inspiral and its amplitude and phase \cite{LackeyKyutoku2014}. More macroscopic and thermal properties of NS can be investigated to verify the examined EOS such as the stability of the $\beta$-equillibrated dense NM relative to the thermodynamic equilibrium \cite{LattimerPrakash2007,Kubis2007,WorleyKrastev2008}, the core-crust transition density and pressure \cite{SeifHashem2022,AttaBasu2014}, the moment of inertia and its crustal fraction for rotating NSs \cite{WorleyKrastev2008,LattimerSchutz2005}, as well as the surface red shift \cite{DouchinHaensel2001}. \par 
In this direction, the LIGO and Virgo collaborations have estimated the radii of the NS binary constituents to be between 10.5 km and 13.3 km \cite{Abbottetal2018}, for masses between 1.16 and 1.60 $\text{M}_{\odot}$. Using different forms of nuclear interaction and many-body approaches, an upper limit of 13.7 km had been deduced for canonical NS radius \cite{AnnalaGorda2018,Fattoyev2018,MostWeih2018,MalikAlam2018,LimHolt2018,KrastevLi2019}. Based on observed data from various NS and pulsars with indicated precise masses between 0.9 and 2.3 $\text{M}_{\odot}$, anticipated radii from 9 to 14 km have been inferred \cite{DouchinHaensel2001,OzelFreire2016,Arzoumanian2018}. Assuming low-spin prior, the dimensionless tidal deformability of canonical NS has been restricted in several studies to be within the limits $\tilde{\varLambda }=300_{-230}^{+420}$ \cite{Abbottetal2019}, ${{\varLambda }_{1.4}}=190_{-120}^{+390}$ \cite{Abbottetal2018}, and its upper limit to be 800 \cite{MalikAlam2018}, at 90\% credibility, whereas limits of $\tilde{\varLambda }\le 700$ \cite{Abbottetal2017} and ${{\varLambda }_{1.4}}\le1400$ \cite{Abbottetal2017} were inferred for the high-spin case, in addition to an upper limit of 1000 \cite{SteinerGandolfi2015} has been estimated from NS X-ray emissions.  Based on a generic family of EOSs for NS matter and results at different baryon density, a lower limit was indicated to be ${{\varLambda }_{1.4}}>120$ \cite{AnnalaGorda2018}, which lies between the lower limits indicated with (without) phase transition to as ${{\varLambda }_{1.4}}>290\,(35.5)$ \cite{MostWeih2018}. The limits $136<{{\varLambda }_{1.4}}<519$ \cite{LimHolt2018} have been indicated based on energy density functionals obtained from the posterior probability distributions, with 95\% credibility. These are a few examples for the various contains indicated for the tidal deformability. \par 
The diverse results of the indicated tidal deformability mentioned above have motivated us to increase the number of observables for which the tidal deformability is investigated, using trustable EOS. One of the appropriate nuclear EOSs is that based on the semi-realistic finite-range M3Y \textit{NN} interaction, which are originally derived from the G-matrix elements of bare nucleon-nucleon potentials. The effective density dependent M3Y-Reid and M3Y-Paris \textit{NN} interactions in their CDM3Y forms are widely employed to investigate the nuclear structure \cite{KhoaLoan2024, LoanLoc2015,SeifAbdelhady2020}, decays \cite{NiRen2011,SeifAdel2019} and reactions \cite{FurumotoSakuragi2009, Seif2008, GhodsiMahmoodi2007}, as well as the nucleus-nucleus potential \cite{ChienKhoa2018,KhoaPhuc2016,SeifAbdelhady2018}, showing factual accuracy and trust. In a non-relativistic Hartree-Fock (HF) scheme, a wide range of soft and stiff EOSs based on the mentioned M3Y interactions have been successfully used to describe the equilibrium \cite{Seif2012} and symmetry \cite{LoanLoc2015,SeifHashem2023} properties of cold \cite{Seif2011,KhoaOertzen1996} and hot \cite{SeifHashem2022,SeifHashem2021} nuclear matter (NM), as well as $\beta$-stable $npe$ and $npe\mu$ NS matter \cite{AttaBasu2014,LoanTan2011,SeifBasu2014,Mukhopadhyay2018}. The same EOSs have been examined using mass-radius constraints for 15 neutron star objects and pulsars, as obtained from electromagnetic and gravitational wave observables using different theoretical schemes. It was indicated that the EOSs from CDM3Y-230 to CDM3Y-270, which are characterized by saturation incompressibility ${K}_{0}=230-270$ MeV, reproduce successfully most of these NS mass-radius constraint \cite{SeifHashem2024}. The EOSs from CDM3Y-230 to CDM3Y-250 also show consistency with experimental constraints and proper energy and pressure calculations on symmetric and asymmetric NM, and pure neutron matter \cite{SeifHashem2023,Seif2011}. Towards our objective, the next section provides brief description of the adopted scheme, the used EOSs, and the method of calculating the considered NS properties. In Sec. \ref{sec:level3}, the tidal deformability of neutron stars, their moment of inertia, and other related quantities, will be investigated using the EOSs based on the semi-realistic M3Y-Paris and Reid \textit{NN} interactions. Finally, a summary of the main results and conclusions are drawn in Sec. \ref{sec:level4}.  
 
\section{\label{sec:level2}THEORETICAL FORMALISM}

Any two neighboring astrophysical objects may get deformed due to expected gravitational tides. The investigation of tidal effects in binary neutron-star (NS) systems could reveal important insights into the EOS of NS matter, as the deformation of a body depends on its internal structure. Gravitational waves (GW) are released when two stars orbit around each other, with anticipated transfer of energy and angular momentum. The orbital frequency increases for shorter orbital radius. The mutual tidal forces between two orbiting NSs deform both of them, which cause the companion star to experience tidal deformability with an induced quadrupole moment ${Q}_{ij}$. Based on Newton theory of gravity, one can define the tidal momentum that drives the quadrupole tidal field in terms of the external gravitational potential ($\Phi$) as \cite{SabatucciBenhar2020},
\begin{equation}
	\label{eqn:EijTm}
	{{E}_{ij}}=-{{\left. \frac{{{\partial }^{2}}\Phi }{\partial {{x}_{i}}\partial {{x}_{j}}} \right|}_{\vec{x}={{{\vec{r}}}_{c}}}},
\end{equation}
The related quadrupole moment can be calculated by,
\begin{equation}
	\label{eqn:Qijqm}
	{{Q}_{ij}}=\frac{1}{3}\int{{{d}^{3}}x\left( 3{{x}_{i}}{{x}_{j}}-{{\delta }_{ij}}{{r}^{2}} \right)}\,{{d}_{m}}(\vec{x}),
\end{equation}
Here, ${\vec{r}}_{c}$ indicates the center of mass of the deforming object exposed to tidal momentum, ${d}_{m}$ denotes the mass density, and ${r}^{2}={\delta }_{ij}{{x}_{i}}{{x}_{j}}$. ${E}_{ij}$ and ${Q}_{ij}$ are traceless symmetric tensors, which are linked under the weak field approximation by the star's quadrupole polarizability or the tidal deformability ($\lambda$), which describes how the star responds to the mutual gravitational field \cite{HindererLackey2010,FlanaganHinderer2008,DamourNagar2009,Hinderer2008,DamourNagar2012}, as
\begin{equation}
	\label{eqn:QM}
	{{Q}_{ij}}=-\lambda \,{{E}_{ij}}.
\end{equation}
In view of General Relativity, $\lambda$ can be defined in terms of the dimensionless so called tidal Love number (${k}_{2}$), the NS radius ($R$), and the gravitational constant ($G$) as \cite{FlanaganHinderer2008,Thorne1998},
\begin{equation}
	\label{eqn:TDF}
	\lambda =\frac{2\,{{k}_{2}}\,{{R}^{5}}}{3\,G}.
\end{equation}
The value of the Love number lies within the range from 0.05 to 0.15 \cite{HindererLackey2010,Hinderer2008,PostnikovPrakash2010}, depending on the star structure. It can be determined in terms of the dimensionless compactness parameter of the star, $\mathpzc{C} =GM/R{c}^{2}$ \cite{MalikAlam2018,PegiosKoliogiannis2022} by
\begin{eqnarray}
	\label{eqn:Lovenu}
	{{k}_{2}}(\mathpzc{C}  ,{{y}_{R}})&=&\frac{8}{5}{{\mathpzc{C}  }^{5}}{{(1-2\mathpzc{C}  )}^{2}}[2-{{y}_{R}}+2\mathpzc{C}  \,({{y}_{R}}-1)]
	\{2\mathpzc{C}  \,[6-3{{y}_{R}}+3\mathpzc{C}  \,(5{{y}_{R}}-8)]\nonumber\\
	&+&4{{\mathpzc{C}  }^{3}}[13-11{{y}_{R}}+\mathpzc{C}  \,(3{{y}_{R}}-2)+2{{\mathpzc{C}  }^{2}}(1+{{y}_{R}})]
	+3{{(1-2\mathpzc{C}  )}^{2}}\nonumber\\&\times&[2-{{y}_{R}}+2\mathpzc{C} ({{y}_{R}}-1)]
	\ln (1-2\mathpzc{C}  ){{\}}^{-1}}.  
\end{eqnarray}
The auxiliary variable ${y}_{R}=y(R)$ is the solution of the first order ordinary differential equation \cite{MalikAlam2018,PegiosKoliogiannis2022},
\begin{equation}
	\label{eqn:ODEy}
	r\frac{dy(r)}{dr}+{{y}^{2}}(r)+y(r)F(r)+{{r}^{2}}Q(r)=0,
\end{equation}
for which the initial condition $y(r=0)=2$ is applied. The functions $F(r)$ and $Q(r)$ are defined in terms of the energy density $\varepsilon(r)$, pressure $P(r)$, and a metric function ${{\left( 1-\frac{2GM(r)}{r{{c}^{2}}} \right)}^{-1}}$ of spherical NS \cite{Glendenning2000}, as \cite{MalikAlam2018,PegiosKoliogiannis2022}
\begin{equation}
	\label{eqn:For}
	F(r)=\left\{ 1-\frac{4\pi {{r}^{2}}G}{{{c}^{4}}}\left[ \varepsilon (r)-P(r) \right] \right\}
	{{\left( 1-\frac{2GM(r)}{r{{c}^{2}}} \right)}^{-1}},
\end{equation}
and
\begin{eqnarray}
	\label{eqn:Qor}
	Q(r)&=&\frac{4\pi G}{{{c}^{4}}}\left[ 5\varepsilon (r)+9P(r)+\frac{\varepsilon (r)+P(r)}{{{({{v}_{s}}/c)}^{2}}} \right]
	{{\left( 1-\frac{2GM(r)}{r{{c}^{2}}} \right)}^{-1}}-\frac{6}{{{r}^{2}}}{{\left( 1-\frac{2GM(r)}{r{{c}^{2}}} \right)}^{-1}} \nonumber\\
	&-&\frac{4{{G}^{2}}{{M}^{2}}(r)}{{{r}^{4}}{{c}^{4}}}{{\left( 1+\frac{4\pi {{r}^{3}}P(r)}{M(r){{c}^{2}}} \right)}^{2}}
	{{\left( 1-\frac{2GM(r)}{r{{c}^{2}}} \right)}^{-2}}.
\end{eqnarray}
The subluminal adiabatic speed of sound ${{v}_{s}}=\sqrt{\partial P(r)/\partial \varepsilon (r)}\le c$ is limited in dense staller matter at zero frequency by the speed of light ($c$), according to the causality condition \cite{DemorestPennucci2010,BaldoFerreira1999}. Starting from the center of the NS, we can employ the fourth order Runge-Kutta technique to numerically solve Eq. (\ref{eqn:ODEy}) self-consistently along with the well-known Tolman-Oppenheimer Volkoff (TOV) equations,
\begin{eqnarray}
	\label{eqn:TOV}
	\frac{dP(r)}{dr}&=&-\frac{G\,\varepsilon (r)\,M(r)}{{{c}^{2}}{{r}^{2}}}\left( 1+\frac{P(r)}{\varepsilon (r)} \right)
	\left( 1+\frac{4\,\pi \,{{r}^{3}}P(r)}{{{c}^{2}}M(r)} \right){{\left( 1-\frac{2\,GM(r)}{r\,{{c}^{2}}} \right)}^{-1}}, \nonumber\\
	\frac{dM(r)}{dr}&=&\frac{4\,\pi \,{{r}^{2}}\varepsilon (r)}{{{c}^{2}}}.
\end{eqnarray}
Here, $r$ and $M(r)$ respectively denote the radial coordinate measured from the center of NS and the gravitational mass enclosed in it. The number of baryons $a(r)$ within a sphere of radius $r$ may be estimated from the TOV equations as
\begin{equation}
	\label{eqn:aor}
	\frac{da(r)}{dr}=\frac{4\,\pi \,{{r}^{2}}\rho (r)}{\sqrt{1-\frac{2\,G\,M(r)}{r\,{{c}^{2}}}}}.
\end{equation}
The baryon number density is defined by $\rho(r)$ in ${\text{fm}}^{-3}$. The radius ($R$), total gravitational mass ${M}_{G}=M(R)$, the auxiliary parameter ${y}_{R}=y(R)$, the total baryon number $A=a(R)$, and the corresponding total baryonic mass of the NS are obtained by the vanishing condition of pressure at the surface ($r=R$). The central pressure ${P}_{c}$ and density ${\rho}_{c}$ are defined at $r=0$, while $a(0)=0$. The dimensionless tidal deformability ($\varLambda$) is then obtained in terms of the Love number and the compactness parameter as \cite{FlanaganHinderer2008,Thorne1998,PegiosKoliogiannis2020},
\begin{equation}
	\label{eqn:DIMTDF}
	\varLambda =\frac{2\,{{k}_{2}}}{3\,{{\mathpzc{C}  }^{5}}}=\frac{2}{3}{{k}_{2}}{{\left( \frac{R\,{{c}^{2}}}{G\,M} \right)}^{5}}.
\end{equation}  
For cold uniform $npe\mu$ matter in $\beta$ equilibrium inside NS core, the total energy density $\varepsilon(r,\rho)$ and pressure can be obtained by adding its baryons ${\varepsilon}_{b}(p,n)$ and leptons ${\varepsilon}_{e,\mu}$ contributions \cite{XuChen2009,Seif2011,SeifHashem2024},
\begin{equation}
	\label{eqn:epstot}
	\varepsilon (\rho ,{{x}_{p}},{{x}_{e}},{{x}_{\mu }})={{\varepsilon }_{b}}(\rho ,{{x}_{p}})+\sum\limits_{\ell =e,\mu }{{{\varepsilon }_{\ell }}(\rho ,{{x}_{\ell }})},
\end{equation}
and
\begin{equation}
	\label{eqn:Ptot}
	P(\rho ,{{x}_{p}},{{x}_{e}},{{x}_{\mu }})={{P }_{b}}(\rho ,{{x}_{p}})+\sum\limits_{\ell =e,\mu }{{{P }_{\ell }}(\rho ,{{x}_{\ell }})},
\end{equation}  
where ${{P}_{b}}(\rho ,{{x}_{p}})={{\rho }^{2}}\partial {(\frac{E}{A}(\rho ,{{x}_{p}}))}/\partial \rho$ and ${{P}_{\ell }}(\rho ,{{x}_{\ell }})={{\mu }_{\ell }}\,\rho \,{{x}_{\ell }}-{{\varepsilon }_{\ell }}(\rho ,{{x}_{\ell }})$. Here, ${x}_{i=p,e,\mu}$ denote the proton, electron, and muon fractions and the total baryon energy can be obtained as
\begin{equation*}
	{{\varepsilon }_{b}}(\rho ,{{x}_{p}})=\rho \,\left[ \frac{E}{A}(\rho ,{{x}_{p}})+{{x}_{p}}\,{{m}_{p}}{{c}^{2}}+(1-{{x}_{p}})\,{{m}_{n}}{{c}^{2}} \right].
\end{equation*}
Based on the CDM3Y density dependent form \cite{KhoaSatchler1997} of the M3Y-Paris \cite{AnantaramanToki1983} and M3Y-Reid \cite{BertschBorysowicz1977} \textit{NN} interactions, the energy per nucleon (${E}/{A}$) reads \cite{Seif2011,SeifHashem2023},  
\begin{eqnarray}
	\label{eqn:EDA}
	\frac{E}{A}\left(\rho ,{{x}_{p}}\right)&=&\frac{3\,{{\hbar }^{2}}\,k_{F}^{2}\left[ {{(2-2{{x}_{p}})}^{5/3}}+{{(2{{x}_{p}})}^{5/3}} \right]}{20\,m}
	+\frac{\rho }{2}\,\left\{ {{F}_{0}}(\rho )\,J_{00}^{D}+{{(1-2{{x}_{p}})}^{2}}\,{{F}_{1}}(\rho )\,J_{01}^{D} \right\} \nonumber\\
	&+&\frac{\rho }{8}\int{d\vec{r}\left[ {{F}_{0}}(\rho )v_{00}^{Ex}{{({{B}_{0}})}^{2}}+{{F}_{1}}(\rho )v_{01}^{Ex}{{({{B}_{1}})}^{2}} \right]}, \nonumber\\
	{{B}_{0}}&=&(2-2\,{{x}_{p}})\,{{{\hat{J}}}_{1}}({{k}_{Fn}}r)+2\,{{x}_{p}}\,{{{\hat{J}}}_{1}}({{k}_{Fp}}r), \nonumber\\
	{{B}_{1}}&=&(2-2\,{{x}_{p}})\,{{{\hat{J}}}_{1}}({{k}_{Fn}}r)-2\,{{x}_{p}}\,{{{\hat{J}}}_{1}}({{k}_{Fp}}r).
\end{eqnarray}
While the leading term in Eq. (\ref{eqn:EDA}) represents the kinetic energy contribution, we have $J_{00(01)}^{D}=\int{d\vec{r}\,v_{00(01)}^{D}(r)}$ where $v_{00(01)}^{D}(r)$ represent the central isoscalar (00) and isovector (01) components of the direct part of the M3Y-Reid \cite{BertschBorysowicz1977} and Paris \cite{AnantaramanToki1983} \textit{NN} interactions \cite{SatchlerLove1979}, and $v_{00(01)}^{Ex}(r)$ are the corresponding exchange components. ${{\hat{J}}_{1}}(x)$ is defined by the first-order spherical Bessel function as ${{\hat{J}}_{1}}(x)=3{{j}_{1}}(x)/x$. The neutron (proton) and total Fermi momenta are respectively defined as ${{k}_{Fn(p)}}={{(3{{\pi }^{2}}{{\rho }_{n(p)}})}^{1/3}}$ and ${{k}_{F}}={{(3\,{{\pi }^{2}}\rho /2)}^{1/3}}$. The isoscalar (${F}_{0}$) and isovector (${F}_{1}$) density dependencies have been added to the bare M3Y \textit{NN} interaction \cite{KhoaSatchler1997} in order to describe the corresponding effective interactions, 
\begin{equation}
	\label{eqn:DensDep}
	\begin{aligned}
		& v_{00(01)}^{D,Ex}(\rho ,r)={{F}_{0(1)}}(\rho )\,v_{00(01)}^{D,Ex}(r), \\ 
		& {{F}_{0(1)}}(\rho )={{C}_{0(1)}}\left( 1+{\alpha}_{0(1)}\,{{e}^{-{\beta}_{0(1)}\,\rho }}-{\gamma}_{0(1)}\,\rho  \right). \\ 
	\end{aligned}
\end{equation}
A variety of CDM3Y-${K}_{0}$ parameterizations (${C}_{i=0,1}$, ${\alpha}_{i}$, ${\beta}_{i}$ and ${\gamma}_{i}$) have been obtained for the isoscalar \cite{Seif2011} and isovector \cite{SeifHashem2023,SeifHashem2024} density dependencies, to describe a wide range of soft and stiff EOSs with saturation incompressibility values ranging from ${K}_{0}$ = 150 MeV to 330 MeV. While the isoscalar density dependence was parameterized to establish the equilibrium properties of symmetric NM using analytical formulas \cite{Seif2011} in non-relativistic HF scheme, the isovector parameters were determined \cite{SeifHashem2023} by iteratively fitting \cite{KhoaThan2007} the  obtained isovector single-nucleon potential, within certain energy and density ranges, with the Brueckner-Hartree–Fock findings reported by earlier studies \cite{JeukenneLejeune1977,Lejeune1980}. On the other hand, the energy density of electrons ($e$) and muons ($\mu$) can be expressed using the non-interacting Fermi gas model \cite{XuChen2009,SeifHashem2024,Tolman1939,OppenheimerVolkoff1939} as,
\begin{eqnarray}
	\label{eqn:epslep}
	{{\varepsilon }_{\ell =e(\mu )}}(\rho ,{{x}_{\ell }})&=&\frac{{{m}_{\ell }}\,{{c}^{2}}}{8\,{{\pi }^{2}}\,\lambda _{\ell }^{3}}[\lambda _{\ell }{{k}_{F\ell }}(\rho ,{{x}_{\ell }})
	 \sqrt{1+{{({{\lambda }_{\ell }}\,{{k}_{F\ell }}(\rho ,{{x}_{\ell }}))}^{2}}} ( 1+2\,{{({{\lambda }_{\ell }}\,{{k}_{F\ell }}(\rho ,{{x}_{\ell }}))}^{2}} )  \nonumber\\
	&-&\ln ( {{\lambda }_{\ell }}\,{{k}_{F\ell }}(\rho ,{{x}_{\ell }})+\sqrt{1+{{({{\lambda }_{\ell }}\,{{k}_{F\ell }}(\rho ,{{x}_{\ell }}))}^{2}}} )],
\end{eqnarray}
where ${{\lambda }_{\ell }}=\hbar c/{{m}_{\ell }}{{c}^{2}}$ and ${{k}_{F\ell}}={{(3{{\pi }^{2}}{{\rho }_{\ell}})}^{1/3}}$ are the leptonic Compton wavelength and Fermi momentum, respectively. The corresponding chemical potential can be written in terms of the rest mass energy (${{m}_{\ell }}{{c}^{2}}$) and ${k}_{F\ell}$ as,
\begin{equation}
	\label{eqn:mul}
	{{\mu }_{\ell }}(\rho ,{{x}_{\ell }})=\sqrt{{{\hbar }^{2}}{{c}^{2}}{{({{k}_{F\ell }}(\rho ,{{x}_{\ell }}))}^{2}}+m_{\ell }^{2}{{c}^{4}}}.
\end{equation}    
The density at which muons begin to emerge is somewhat lower than the saturation nuclear density \cite{LoanTan2011,SeifHashem2024}. An electron chemical potential larger than the muon rest mass energy is required for effective muon contribution to chemical equilibrium. In $\beta$-stable $npe\mu$ matter, the direct URCA reactions correlate with the chemical equilibrium condition \cite{Seif2011,SeifHashem2024},
\begin{equation}
	\label{eqn:equill}
	{{\mu }_{e}}={{\mu }_{\mu }}={{\mu }_{n}}-{{\mu }_{p}}=-\frac{\partial{(\frac{E}{A}(\rho ,{{x}_{p}}))}}{\partial {{x}_{p}}}.
\end{equation}
Using Eqs. (\ref{eqn:mul}) and (\ref{eqn:equill}) and the charge neutrality condition (${x}_{p}={x}_{e}+{x}_{\mu}$), the $p$, $e$, and $\mu$ fractions can be determined using the obtained relation \cite{LoanTan2011,SeifHashem2022}, 
\begin{equation}
	\label{eqn:pemufrac}
	3{{\pi }^{2}}{{(\hbar c)}^{3}}\rho {{x}_{p}}-\mu _{e}^{3}-{{\left[ \mu _{e}^{2}-{{({{m}_{\mu }}{{c}^{2}})}^{2}} \right]}^{3/2}}\Theta ( {{\mu }_{e}}-{{m}_{\mu }}{{c}^{2}})=0,
\end{equation}
In this equation, $\Theta$ represents Heaviside step function. Now, the transition density (${\rho}_{t}$), pressure (${P}_{t}$), and proton fraction (${x}_{pt}$) at the inner edge at which the solid crust appears next to the liquid NS core can be obtained using the positivity condition of compressibility (${K}_{{{\mu }_{i}}}$) at a certain chemical potential (${\mu}_{i}$) \cite{SeifBasu2014,Kubis2007},
\begin{equation}
	\label{eqn:kmu}
	{{K}_{{{\mu }_{i}}}}={{\left( \frac{\partial P}{\partial \rho } \right)}_{{{\mu }_{i}}}}=\frac{K(\rho ,{x}_{p})}{9}-\frac{{{\left( \rho\, \frac{{{\partial }^{2}}{({E}/{A})}}{\partial \rho \partial {x}_{p}} \right)}^{2}}}{\frac{{{\partial }^{2}}{({E}/{A})}}{\partial {{{x}_{p}}^{2}}}}>0,
\end{equation}
The second term in Eq. (\ref{eqn:kmu}) originates from the lepton pressure. The incompressibility coefficient of asymmetric nuclear matter (ANM) of a given proton fraction (${x}_{p}$) is given by \cite{Seif2011,SeifHashem2023,SeifHashem2024}
\begin{equation}
	\label{eqn:Kb}
	K(\rho ,{x}_{p})=9\left( 2\rho \,\frac{\partial {({E}/{A})}}{\partial \rho }+{{\rho }^{2}}\,\frac{{{\partial }^{2}}{({E}/{A})}}{\partial {{\rho }^{2}}} \right).
\end{equation}

In the present calculations, the EOS derived by Haensel–Zdunik \cite{Haensel1989} and Feynman-Metropolis-Teller \cite{Feynman1949} will be used to describe the outer crust of the NS, while the inner crust will be described by the EOS given by Douchen \textit{et al}. \cite{DouchinHaensel2001,Douchin2000}.

\section{\label{sec:level3}RESULTS AND DISCUSSION}
Tidal effects in a binary system can be scaled by the corresponding dimensionless quadrupole deformability of the participating NS. Important information related to the structure of NSs can be indicated by the relationship between the tidal deformability and the NS radius ($R$) and the corresponding second love number (${k}_{2}$). In this section, we use experimentally indicated constraints on the tidal deformability of NS objects, which obtained from analyzing recent GW and pulsar observations in different theoretical schemes, to examine the reliability and validity of the EOS derived from the semi-realistic M3Y-Paris and Reid \textit{NN} interactions to describe asymmetric nuclear matter (ANM) up to extremely high density. We consider different CDM3Y-${K}_{0}$ forms of the EOS with saturation incompressibility values ranging from ${K}_{0}=200$ MeV to 330 MeV, to describe soft and stiff ANM, with isoscalar and isovector density dependence parameters of Eq. (\ref{eqn:DensDep}) are given in Refs. \cite{SeifHashem2023, SeifHashem2024}. These EOSs were obtained using non-relativistic Hartree–Fock scheme \cite{Seif2011,SeifHashem2023}, and they have been successfully employed to give the mass-radius profiles of non-rotating NS by solving the TOV equations at hydrostatic equilibrium \cite{SeifHashem2024}. The stiffness of the EOS is also specified by the symmetry energy coefficient ($E_{sym2}$) and the density-slope of the symmetry-energy ($L$). At the saturation-density $\rho_0$ of SNM, $E_{sym2}$ and $L$ have values of 29.10 (30.98) MeV and 48.18 (50.81), respectively, for the CDM3Y-Paris (Reid) EOSs. For a given isospin-asymmetry, $L$ linearly increases with $K_0$, while its value at the equilibrium density of ANM of a given isospin-asymmetry decreases with increasing its isospin-asymmetry \cite{HashemHassanien2024}. Meanwhile, the central isospin-asymmetry of the NS matter at $\beta$-equilibrium increases with $K_0$. These three factors combine to indicate the change of the density-slope of the symmetry energy with $K_0$ of the employed EOS, at $\rho_{0I}$ of the central isospin-asymmetry ($I_c$) of $\beta$-stable matter of NS at its core center. For instance, $I_c$ inside NS (1.4 $\text{M}_{\odot}$) increases from 0.35 to about 0.84 with increasing $K_0$ (CDM3Y-Paris) from 200 MeV to 330 MeV. This consequently deceases the density-slope, at $\rho_{0I}$($I_c$), from 44.71 MeV to 35.68 MeV, respectively. The corresponding symmetry-energy coefficient decreases from 27.72 MeV to 22.07 MeV. For NS ($\text{M}_{\odot}$), the three mentioned factors combine to yield almost unchanged $L$($\rho_{0I}$).  \par

\begin{figure*}[!htbp]
	\centering
	\subfigure[ \label{fig:Fig1a}]{
		\includegraphics[height=12cm,width=1.0\linewidth]{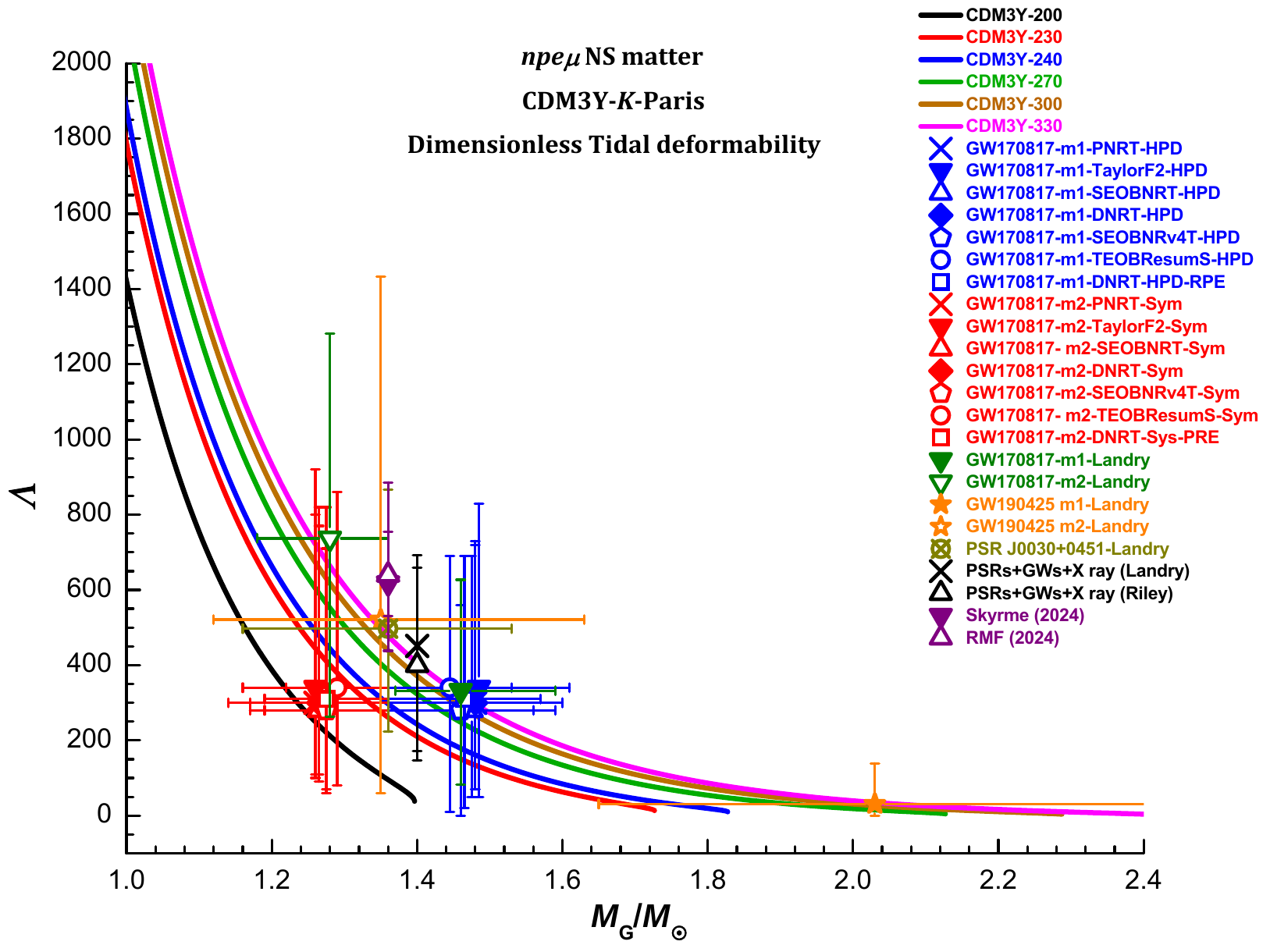}}
	~
	\centering
	\subfigure[ \label{fig:Fig1b}]{
		\includegraphics[height=12cm,width=1.0\linewidth]{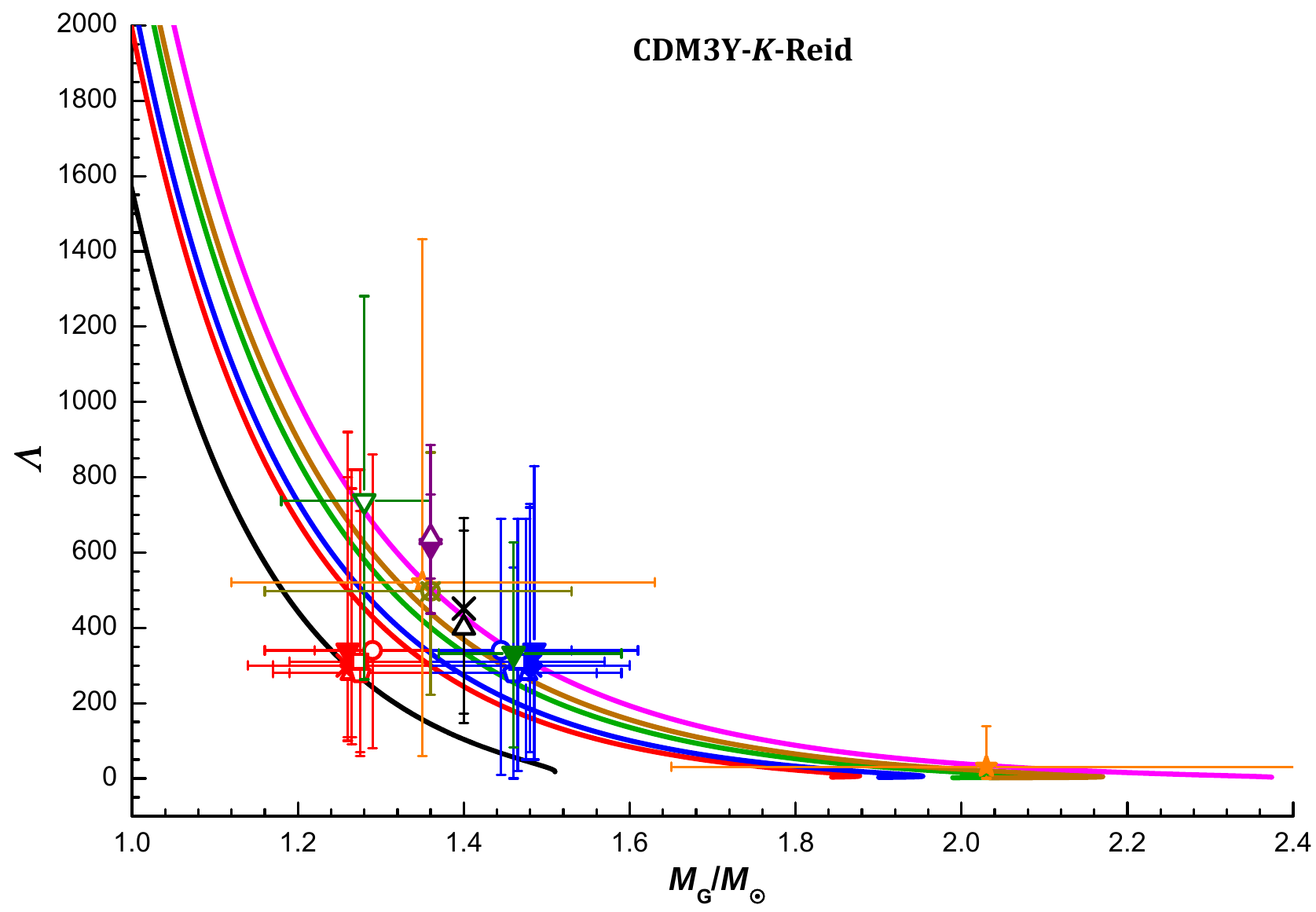}}	
\end{figure*}

\begin{figure*}[!htbp]
	\subfigure[ \label{fig:Fig1c}]{
		\includegraphics[height=13cm,width=1.0\linewidth]{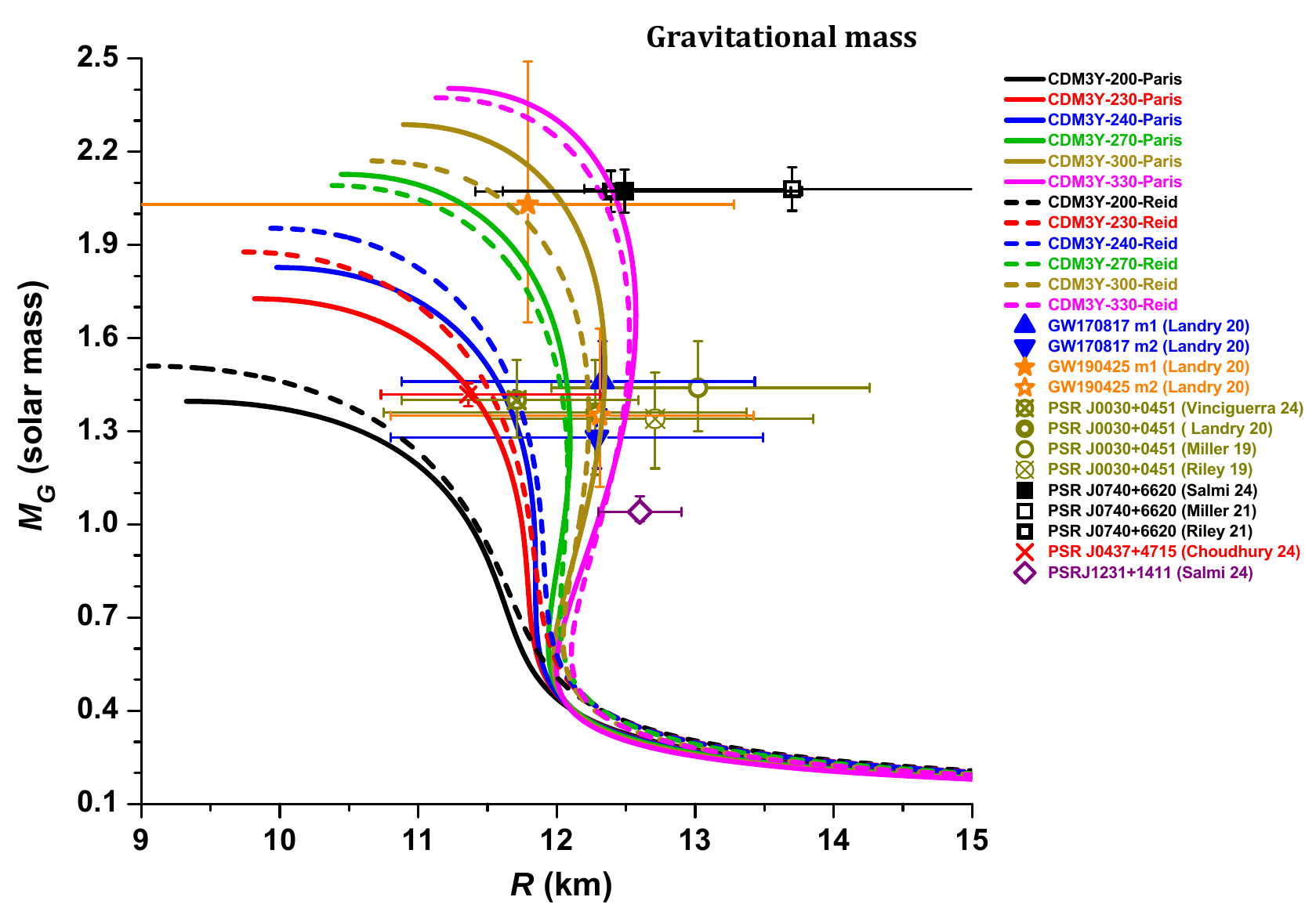}}	
	\caption{\label{fig:Figure1} Tidal deformability versus the NS gravitational mass, based on (a) CDM3Y-Paris and (b) CDM3Y-Reid hadronic EOSs, which describe soft and stiff NM of saturation incompressibility coefficient 200 MeV$\leq$ ${K}_{0}$ (SNM)$\leq$ 330 MeV. To evaluate the present results, we added experimentally indicated values \cite{Abbottetal2019} for primary (\textit{$m_1$}) and secondary (\textit{$m_2$}) partners of NS binaries from GW170817 \cite{Abbottetal2017,Abbottetal2019} and GW190425 \cite{Abbottetal892L32020} GW events, 1.36 $\text{M}_{\odot}$ and 1.40 $\text{M}_{\odot}$ NSs \cite{JieSedrakian2023,ThomasRiley2021}, and pulsars (PSRs), in addition to recent calculations based on Skyrme and RMF EOSs \cite{AdilImam2024}. (c) Gravitational NS mass versus its radius, based on the CDM3Y-$K$ EOSs in (a) and (b), with the NS mass-radius empirical data inferred for the PSR J0030+0451, J0740+6620, J0437+4715, and J1231+1411 NS \cite{ThomasRiley2021,ThomasRiley2021,Salmietal2024a, Choudhuryetal2024, Salmietal2024b, Vinciguerraetal2024, Milleretal2021,Milleretal2019} X-ray pulsars from NICER data, and that inferred for the gravitational wave events GW170817 and GW190425 \cite{LandryEssick2020}.}  
\end{figure*}

In the beginning, we specifically focus on the influence of the employed EOS on the estimated tidal response of the NS. While the inner and outer regions of the NS core are assumed to be in the form of $npe\mu$ matter at $\beta$-equilibrium, the crustal region probably contain bound clusters and finite nuclei with mass numbers up to $A=100$. Figure \ref{fig:Figure1} shows the dimensionless tidal deformability of a given non-rotating spherically symmetric NS as a function of the gravitational stellar mass $M\,(\text{M}_{\odot})$, based on the CDM3Y-$K$ Paris (in Fig. \ref{fig:Fig1a}) and Reid (in Fig. \ref{fig:Fig1b}) EOSs characterized by ${K}_{0}=200-330$  MeV. We added in Figs. \ref{fig:Fig1a} and \ref{fig:Fig1b} indicated constraints on the tidal deformability inferred from analyzing the observed gravitational wave events GW170817 \cite{Abbottetal2017} and GW190425 \cite{Abbottetal892L32020}, from the coalescence of the primary (\textit{m1}) and secondary (\textit{m2}) components of NS binaries. To obtain these constraints \cite{Abbottetal2019}, the detected GW data have been analyzed based on the PhenomPNRT, PhenomDNRT, TaylorF2, SEOBNRT, SEOBNRv4T, and TEOBResumS waveform models, with LALINFERENCE and RAPIDPE (RPE) codes \cite{Abbottetal2019}. In these analyses, both the symmetric (Sym) 90\% credible interval and the corresponding 90\% highest posterior density (HPD) interval have been individually considered, along with restricting the magnitude of the component spins \cite{Abbottetal2019}. The estimations by Landry \textit{et al}. on the GW170817 and GW190425 events and on the millisecond pulsar PSR J0030+0451 in Fig. \ref{fig:Figure1} were obtained at 90\% credible level with non-parametric constraints on the NS matter and its EOS from analysis of combined astronomical data of GW and pulsar observations \cite{LandryEssick2020}. The displayed values of the tidal deformability of NS (1.4 $\text{M}_{\odot}$) in Fig. \ref{fig:Figure1} were inferred by Landry \textit{et al}. \cite{LandryEssick2020} and Riley \textit{et al}. \cite{ThomasRiley2021} by quoting the median based on the joint constraints of PSRs, GWs, and X-ray data. The estimated tidal deformability of NS (1.36 $\text{M}_{\odot}$) that obtained using EOSs based on non-relativistic mean-field calculations in terms of Skyrme effective interactions and based on relativistic mean-field (RMF) model in terms of effective Lagrangian are also added in Figs. \ref{fig:Fig1a} and \ref{fig:Fig1b} for comparison \cite{AdilImam2024}. These median results are obtained considering 95\% confidence interval. Figure \ref{fig:Fig1c} shows the mass-radius profiles of non-rotating NS, based on the EOSs considered in panels (a) and (b) of Fig. \ref{fig:Figure1}, compared with the masses and radii constraints indicated for the PSR J0030+0451, J0740+6620, J0437+4715, and J1231+1411 NS \cite{ThomasRiley2021,ThomasRiley2021,Salmietal2024a, Choudhuryetal2024, Salmietal2024b, Vinciguerraetal2024, Milleretal2021,Milleretal2019}, based on the NICER observations of X-ray pulsars, and those deduced from the gravitational wave events GW170817 and GW190425 \cite{LandryEssick2020}. As seen in Fig. \ref{fig:Fig1c}, the CDM3Y (${K}_{0}\geq$ 230 MeV) equations of state successfully reproduce most of the obtained mass-radius constraints. The NS masses heavier than 2 $\text{M}_{\odot}$ indicate stiffer EOS of ${K}_{0}>$ 240 MeV. The estimated maximum gravitational mass of NS and its corresponding radius $R({M}_{\text{max}})$ increase with increasing the isoscalar incompressibility of the employed EOS. Increasing the symmetry energy coefficient (${E}_{sym2}$) and its corresponding density-slope ($L$) from the CDM3Y-Paris to the CDM3Y-Reid forces leads to a slight rise in the predicted ${M}_{\text{max}}$ and ${R}_{\text{max}}$, with more noticeable increase in their corresponding $R({M}_{\text{max}})$ and $M({R}_{\text{max}})$, and in the indicated tidal deformability $\varLambda ({M}_{\text{max}})$, for a given EOS of of ${K}_{0}>240$ MeV. On the other hand, the gradient of the change in the estimated radius, as a function of mass, decreases for the harder EOS. For instance, a decrease in the NS mass of about 0.5 $\text{M}_{\odot}$ below the indicated ${M}_{\text{max}}$ yields an increase of the corresponding radius of about 1.4 km and 2.4 km for the EOS of $K_0$ = 300 MeV and 200 MeV, respectively. The wide gradient of the mass-radius dependence for the NS lighter than 0.5 $\text{M}_{\odot}$ exhibits weak dependence on $K_0$. This supports the trend seen in Ref. \cite{TanKhoa2021} based on CDM3Y-Paris (${K}_{0}=217-257$ MeV) and BDM3Y (270 MeV) EOSs, and generalizes it across wider spectrum of ${K}_{0}$, from extremely soft to extremely stiff limits. This behavior is further influenced by reducing the negative contributions of the higher-order isovector coefficient of the incompressibility (${K}_{2}$), the isoscalar skewness (${Q}_{0}$), and the kurtosis (${I}_{2}$) symmetry. Specifically, when the ${K}_{0}$ of the CDM3Y-Paris EOS decreases from 300 MeV to 200 MeV, these coefficients shift from -36 MeV, -114 MeV, and -4755 MeV to -17 MeV, -697 MeV, and -5043 MeV, respectively. Meanwhile, the ${Q}_{2}({I}_{0})$ symmetry coefficient increases from 356 MeV (140 MeV) to 468 MeV (2526 MeV). These ${K}_{n}$, ${Q}_{n}$, and ${I}_{n}$ symmetry coefficients respectively describe the curvature, skewness, and kurtosis of the symmetry energy’s density dependence \cite{SeifHashem2025}, where the subscript $n$ denotes the order of isospin asymmetry within the EOS expansion. For the CDM3Y-Paris EOSs, increasing the stiffness of the nuclear matter (NM) reduces the values of the ${I}_{0}$ and ${Q}_{2}$ coefficients while diminishing the negative contributions of both the isoscalar ${Q}_{0}$ and the isovector ${I}_{2}$ coefficients \cite{SeifHashem2025}. Conversely, it enhances the negativity of the ${K}_{2}$ coefficient. These changes lead to larger estimates for the neutron star radius, its tidal deformability, and maximum mass indicated for a given equation of state, as well as an increase in the maximum compactness. However, they also imply a reduction in the compactness of a typical NS and a decrease in the minimum tidal deformability ${\varLambda}_{\text{min}}({M}_{\text{max}})$ \cite{SeifHashem2025}. This trend is evident in Figs. \ref{fig:Fig1a} and \ref{fig:Fig1c} and is further illustrated in Fig. \ref{fig:Fig3a} for the NS compactness.
\par 
Figures \ref{fig:Fig1a} and \ref{fig:Fig1b} show that the tidal deformability is sensitive to the employed EOS, and it reflects its stiffness. For a given stellar mass, the harder NM yields larger deformability and extend the deformability to larger stellar masses, for which the deformability vanishes. For instance, the tidal deformability of a canonical NS (1.4 $\text{M}_{\odot}$) increases from 32 (98) to 413 (428) upon increasing the SNM incompressibility coefficient of the employed CDM3Y-Paris (Reid) EOS from ${K}_{0}=200$ MeV to 330 MeV, respectively, which indicates increasing the rate of inspiral upon increasing $\varLambda$.
This range of $\varLambda$ encompasses the previous estimates based on the CDM3Y-Paris EOSs, $\varLambda$(${K}_{0}=217-257$ MeV) $=142-348$, and that obtained based on the BDM3Y-Paris EOS,  $\varLambda$ (${K}_{0}=270$ MeV) = 407 \cite{TanKhoa2021}. The five EOSs (${K}_{0}=230-330$ MeV) based on both M3Y-Paris and M3Y-Reid \textit{NN} interactions show good agreement with the estimated tidal deformability of the 1.4 $\text{M}_{\odot}$, as indicated by the PSRs+GWs+X ray constraints \cite{ThomasRiley2021,LandryEssick2020}, and with those estimated for NS (1.36 $\text{M}_{\odot}$) based on the Skyrme and RMF calculations \cite{AdilImam2024}. Also, the results based on same EOSs are in overall good agreement with the indicated tidal deformability of the GW170817 and GW190425 events \cite{Abbottetal2019}. The ten EOSs (${K}_{0}=230-330$ MeV) of Paris and Reid interactions likely support the lower limits of the indicated constraints of the heavy-mass component (\textit{m1}) of the GW170817 and GW190425 events and those of the 1.36 $\text{M}_{\odot}$ and 1.4 $\text{M}_{\odot}$ NS masses, but they support the larger indicated constraints on the tidal deformability of lighter masses. The same EOSs (${K}_{0}=230-330$ MeV) fairly verify most of the indicated constraints on the masses and radii of 15 NS objects and pulsars \cite{SeifHashem2024}. Only the soft EOS (${K}_{0}=200$ MeV) fails to reproduce the indicated $\varLambda$ for the NS objects of masses larger than 1.36 $\text{M}_{\odot}$. This is because it indicates small maximum mass of NS, which does not exceed 1.4 $\text{M}_{\odot}$ (1.5 $\text{M}_{\odot}$) for the CDM3Y-Paris (Reid) EOS of ${K}_{0}=200$ MeV \cite{SeifHashem2024}. \par 

As seen in Fig. \ref{fig:Fig1c}, the CDM3Y-$K$ equations of state of ${K}_{0}\leq$ 240 MeV do not support the presence of NS heavier than 2 $\text{M}_{\odot}$. Meanwhile, the CDM3Y-240-Paris  and CDM3Y-(230, 240)-Reid EOSs maintain simultaneous consistency with experimental NM constraints estimated from data on matter flow in relativistic collisions of heavy ions through accurate calculations on energy, symmetry energy, and pressure of both SNM and PNM \cite{SeifHashem2023,DanielewiczLacey2002}. The values of ${K}_{0}=220-240$ MeV commonly lie within the range of isoscalar incompressibility indicated in studies on nuclear structure \cite{XuZhang2021,GargColo2018,Seif2006}. Also, the CDM3Y-($220-240$) \textit{NN} interactions successfully reproduce nuclear structure and reaction data \cite{GhodsiMahmoodi2007,SeifAbdelhady2020b}. This raises the question regarding the existence of hybrid NS stars with many phase transitions \cite{PagliaraSchaffner2008}, including two-flavor color and color-flavor-locked superconducting quark matter states \cite{AlfordSchmitt2008,ZdunikHaensel2013}. Such possible dense matter phase transitions from baryonic matter to quark matter, to hyperons through the exchange of mesons with additional repulsion, or to kaons, would increase the stiffness of the baryon EOS at high density, increasing the anticipated maximum NS mass based on a given EOS \cite{FortinZdunik2015,FortinAvancini2017}. Generally, the CDM3Y-${K}_{0}$($200-300$ MeV) equations of state indicate a pressure band of $\beta$-stable NS matter, as a function of $\rho$ within the ($0.5-1.5$)${\rho}_{0}$ density range, which is lying between the two single polytropic (${P}_{0}({\rho/{\rho}_{0})}^{\Gamma}$) curves given by (${P}_{0},\Gamma$) = (2.443 MeV fm$^{-3}$, 3.293) and (2.449 MeV fm$^{-3}$, 3.904) for CDM3Y-Paris interaction, and given by (2.583 MeV fm$^{-3}$, 3.027) and (2.560 MeV fm$^{-3}$, 3.594) for the CDM3Y-Reid interactions, going from ${K}_{0}$ = 200 MeV to 300 MeV. The obtained pressure bands for the EOSs (${K}_{0}\leq$ 300 MeV) are in consistence with the similar bands obtained based on the prior EOS ensembles given by recent $\chi\text{EFT}$ chiral effective field calculations at N$^2$LO and N$^3$LO next-to-leading orders \cite{Rutherford2024,KellerHebeler2023}, with constraining to recent NICER mass–radius measurements. These pressure bands are described by the curves (${P}_{0},\Gamma$) = ($1.814-3.498$ MeV fm$^{-3}$, $2.391-3.002$) for N$^2$LO, and ($2.207-3.056$ MeV fm$^{-3}$, $2.361-2.814$) for N$^3$LO \cite{Rutherford2024}. The indicated consistency increases with decreasing the stiffness of the CDM3Y-$K$ EOS from that of ${K}_{0}$ = 300 MeV down to 200 MeV. The CDM3Y-330 EOS yields relatively less (higher) pressure at $\rho<{\rho}_{0}$ ($\rho>{\rho}_{0}$), with consistent pressure only within the saturation nuclear density. Some evidences for the isoscalar incompressibility larger than 300 MeV have been indicated in a few studies on giant monopole resonance and NS matter \cite{Sto14,Sto94}. \par

\begin{figure*}[!htbp] 
	\centering	
	\subfigure[ \label{fig:Fig2a}]{
		\includegraphics[height=11.5cm,width=1.0\linewidth]{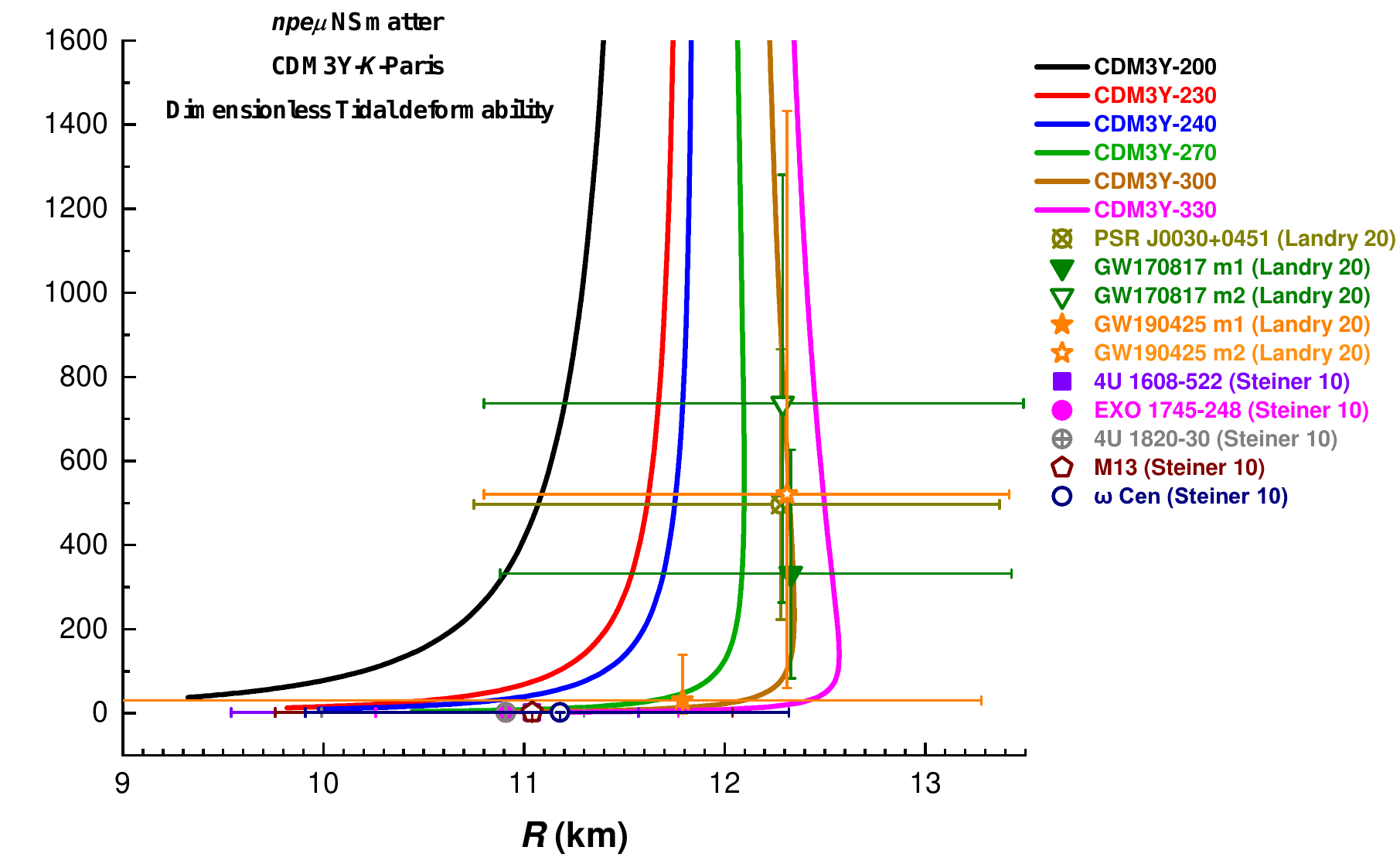}}
	~
	\centering
	\subfigure[ \label{fig:Fig2b}]{
		\includegraphics[height=11.5cm,width=1.0\linewidth]{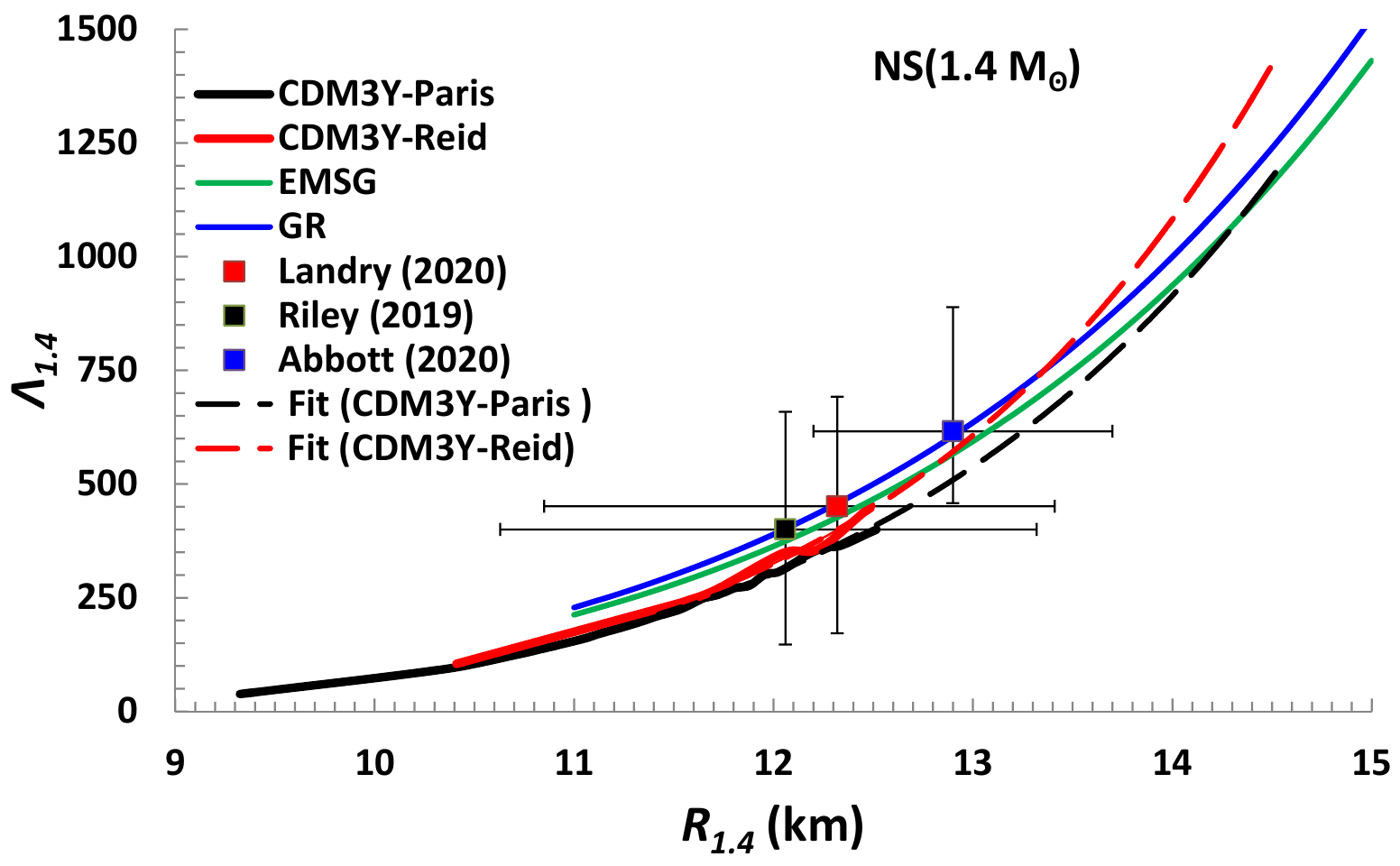}}
	\caption{\label{fig:Figure2} (a) The same as in Fig. \ref{fig:Figure1}(a), but the tidal deformability is plotted versus the NS radius. (b) The obtained radius of NS (1.4 $\text{M}_{\odot}$) based on the CDM3Y-Paris EOSs considered in (a) and the corresponding Reid EOSs, and curve fits to the results based on the CDM3Y-Paris ($\varLambda_{1.4}=5.74\times{10}^{-6}{(R_{1.4}\text{(km)})}^{7.16}$) and Reid ($1.24\times{10}^{-6}{(R_{1.4}\text{(km)})}^{7.80}$) EOSs of ${K}_{0}=230-330$ MeV, compared with similar curve fits for EMSG ($8.37\times{10}^{-5}{(R_{1.4})}^{6.15}$) and GR ($9.67\times{10}^{-5}{(R_{1.4})}^{6.12}$) calculations \cite{AlamPal2024}, and with experimentally inferred constraints by Landry \cite{LandryEssick2020}, Riley \cite{ThomasRiley2019}, and Abbott \cite{Abbottetal896L442020} \textit{et als}.}  
\end{figure*}

\begin{figure*}[!htbp]
	\centering	
	\subfigure[ \label{fig:Fig3a}]{
		\includegraphics[height=12cm,width=1.0\linewidth]{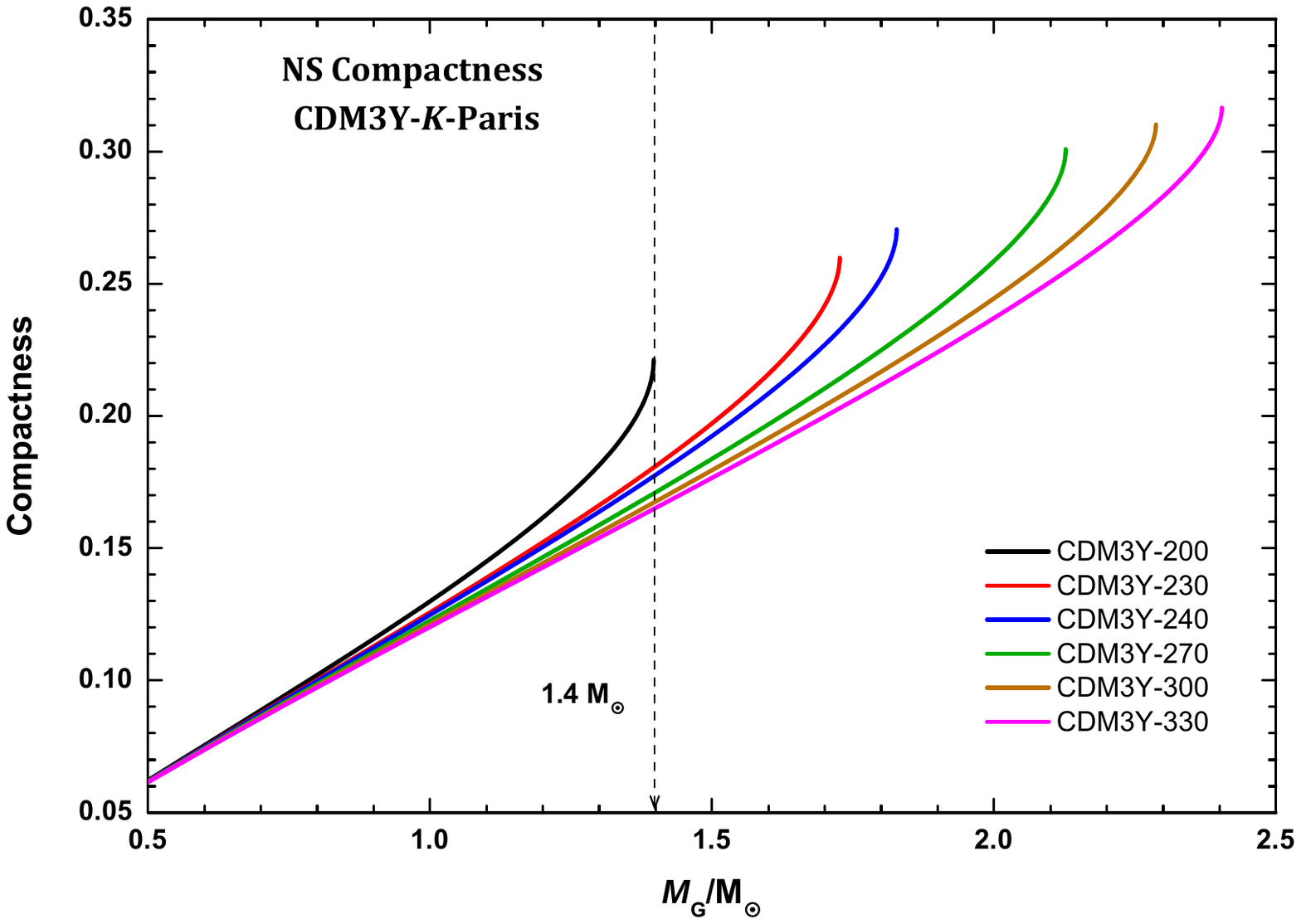}}
	~
	\centering
	\subfigure[ \label{fig:Fig3b}]{
		\includegraphics[height=12cm,width=1.0\linewidth]{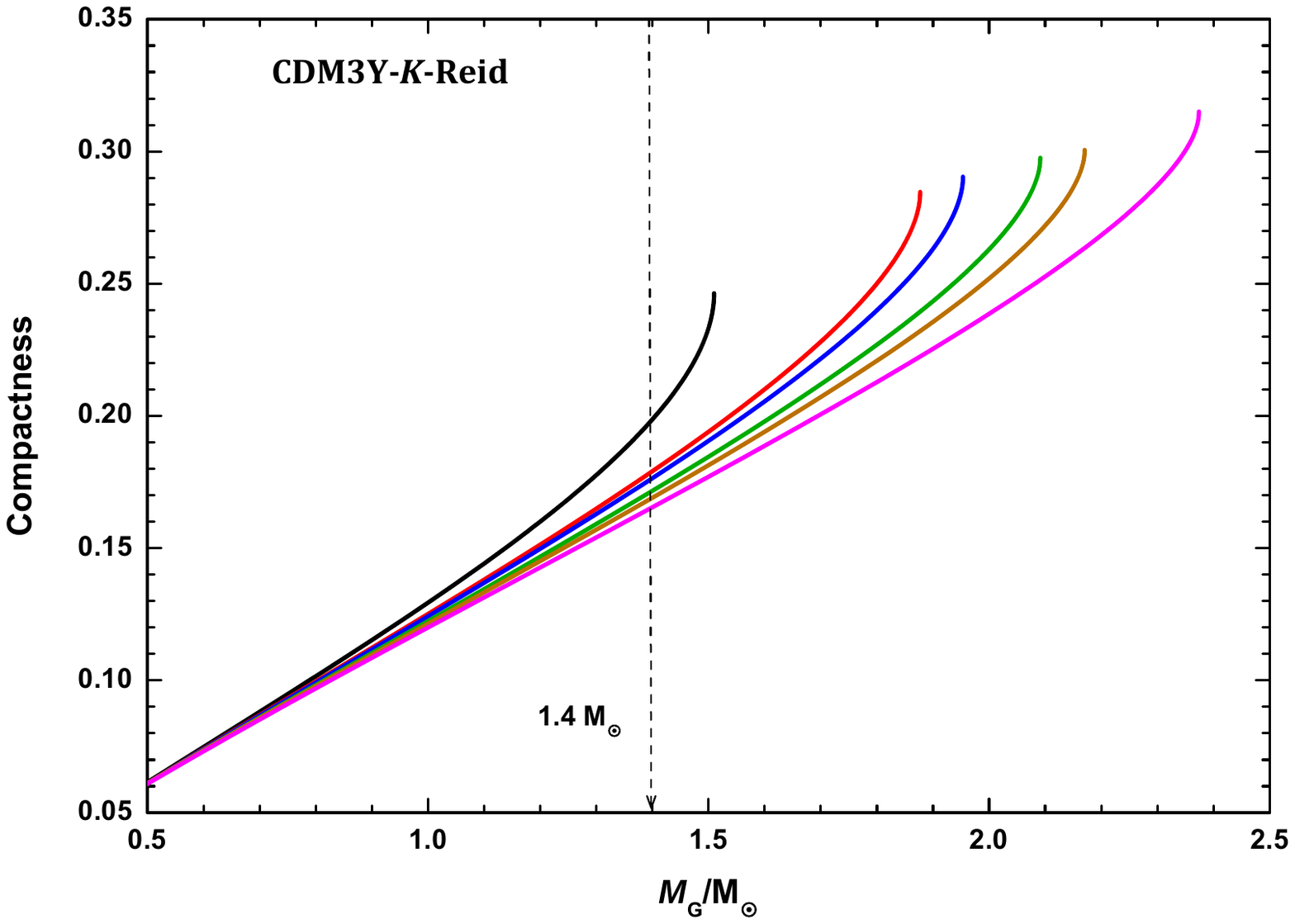}}	
\end{figure*}

In Fig. \ref{fig:Figure2}(a), we again display the tidal deformability of a uniform NS but as a function of its radius (in km), based on the same EOSs in Fig. \ref{fig:Figure1}(a). The obtained calculations are compared in Fig. \ref{fig:Figure2}(a) with the indicated tidal deformability and radii of the 4U 1820-30, 4U 1608-522, EXO 1745-248, $\omega$ Cen, and M13 NS objects \cite{SteinerLattimer2010}, and PSR J0030+0451 pulsar \cite{LandryEssick2020}, in addition to those of the GW170817 and GW190425 \cite{LandryEssick2020} GW events for the primary and secondary binary NS components. The direct proportionality between the radius of the NS and its tidal deformability is understood where $\varLambda$ assess the deviation of the gravitational field of NS relative to a corresponding point mass \cite{HindererLackey2010}. As seen in Fig. \ref{fig:Figure2}(a), the stiffer NM extends the deformability to larger stellar radii, at which the deformability increases sharply. This limiting radius increases from 11.39 km (11.42 km) for the CDM3Y-200-Paris (Reid) EOS to 12.59 (12.54) for the corresponding CDM3Y-330 EOS. Consequently, the stiffer EOS provides larger radius corresponding to a certain deformability of a given NS. These limiting radii correspond to the maximum radii anticipated for NS of mass larger than the solar mass, based on the considered EOS. The increase in the tidal deformability based on the softer EOSs of ${K}_{0}=200-240$ MeV is smoother than that based on the stiffer EOSs because the corresponding estimated radius of the NS of mass less than the solar mass continually increases upon decreasing the mass of the NS, as seen in Fig. \ref{fig:Figure1}(c). The stiffer EOSs indicate a maximum radius for NS ($M>\text{M}_{\odot}$), after which the indicated radius decreases with decreasing the NS mass down to about 0.4 $\text{M}_{\odot}$ then it turns to increase again for the lighter NSs. The indicated ranges of $\Delta R$ within which the estimated tidal deformability multiplies from 100 to 1000 decreases from 1.09 (0.91) km, 0.54 (0.44) km for the CDM3Y-200 and CDM3Y-230 EOSs of Paris (Reid) interactions to 0.13 (0.19) km for the corresponding CDM3Y-270 EOSs. Generally, Figs. \ref{fig:Figure1} and \ref{fig:Figure2} show that the experimental estimates of the tidal deformability can add restrictions on the equation of state, but their uncertainty ranges still quite wide in which a wide range of EOSs of ${K}_{0}$ between 230 and 330 MeV verify them or lay on their boundary. \par 

Figure \ref{fig:Fig2b} displays the radius of a canonical 1.4 $\text{M}_{\odot}$ NS as obtained based on the considered CDM3Y-Paris and Reid EOSs. The obtained results based on the EOSs (${K}_{0}=230-330$ MeV) of CDM3Y-Paris and Reid interactions can be respectively fitted rather accurately to read $\varLambda_{1.4}=5.74\times{10}^{-6}{(R_{1.4}\text{(km)})}^{7.16}$ and $1.24\times{10}^{-6}{(R_{1.4}\text{(km)})}^{7.80}$. Similar expressions \cite{AlamPal2024} obtained from fits to calculations based on energy-momentum squared gravity (EMSG), $8.37\times{10}^{-5}{(R_{1.4})}^{6.15}$, and the general theory of relativity (GR) of Einstein, $9.67\times{10}^{-5}{(R_{1.4})}^{6.12}$, as well as experimentally inferred constraints \cite{JieSedrakian2023,ThomasRiley2019,Abbottetal896L442020}, are presented in Fig. \ref{fig:Fig2b} for comparison. Although such obtained relations show independence of the EOS but they still have some degree of model dependence. \par 

In Table \ref{table1}, we list in its fifth and sixth columns the estimated radius ($R$), in km, and the tidal deformability ($\varLambda$) as obtained based on the CDM3Y-Paris and Reid EOSs characterized by 230 MeV$\leq$ ${K}_{0}$ (SNM)$\leq$ 330 MeV, for the primary and secondary partners of the NS binaries GW170817 and GW190425, which have the experimentally indicated masses given in column 2. The same quantities are also given in Table \ref{table1} for NS that have the maximum mass (${M}_{\text{max}}$) and maximum radius (${R}_{\text{max}}$) indicated by the considered EOSs, and for NS (1.40 $\text{M}_{\odot}$). To evaluate the present calculations, the experimentally inferred values of the radius and tidal deformability, for the GW170817 and GW190425 binaries and for NS (1.40 $\text{M}_{\odot}$), are respectively given in columns 3 and 4. Presented in the last three columns are recent estimates of ${M}_{\text{max}}$, its radius, and the radius and tidal deformability of NS (1.4 $\text{M}_{\odot}$), based on EOSs of given ranges of $K_0$, which are derived from mean-field interactions \cite{SantosMalik2024, KumarThakur2023, MalikPais2024, LiTian2024, ThakurKumar2022a, ThakurKumar2022b, HuangShen2024}. These calculations are performed using a Bayesian inference approach based on RMF description of NM couplings with isovector scalar $\delta$-meson field in Refs. \cite{SantosMalik2024, ThakurKumar2022a}, and with couplings of isoscalar-scalar $\sigma$, isoscalar-vector $\omega$, and isovector-vector $\rho$ meson fields in \cite{KumarThakur2023}, and based on twenty one unified EOSs of NS within non-linear RMF description in \cite{MalikPais2024}, while Ref. \cite{LiTian2024} considers covariant density functionals with density-dependent meson-nucleon coupling. The calculations from Ref. \cite{ThakurKumar2022b} are obtained based on seven EOSs of pure hadronic matter and three hybrid EOSs of superdense nucleon-quark matter, while one of the nine RMF interactions considered in \cite{HuangShen2024} includes tensor coupling. Although the EOSs considered  here are of wide incompressibility range between ${K}_{0}=230$ MeV and 330 MeV, but they yield narrower ranges of the indicated $R$ and $\varLambda$ than their experimentally estimated constraints, and than those indicated by calculations based on more limited ranges of incompressibility. This introduces a step towards improved interpretation of astrophysical and nuclear data \cite{Koe24}. \par 

\begin{figure*}[!htbp]
	\centering	
	\subfigure[ \label{fig:Fig3c}]{
		\includegraphics[height=11.5cm,width=1.0\linewidth]{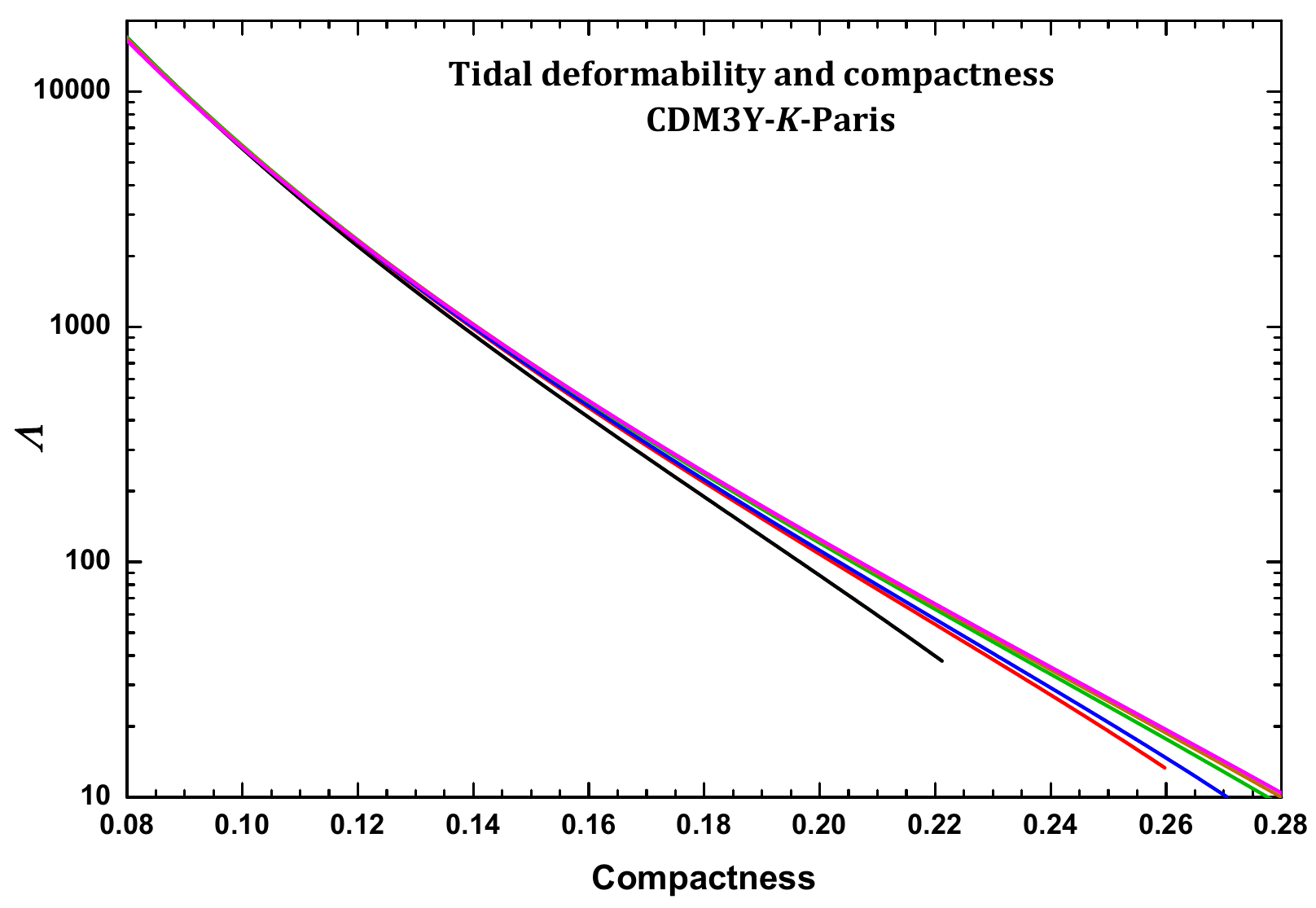}}
	~
   \centering
	\subfigure[ \label{fig:Fig3d}]{
		\includegraphics[height=11.5cm,width=1.0\linewidth]{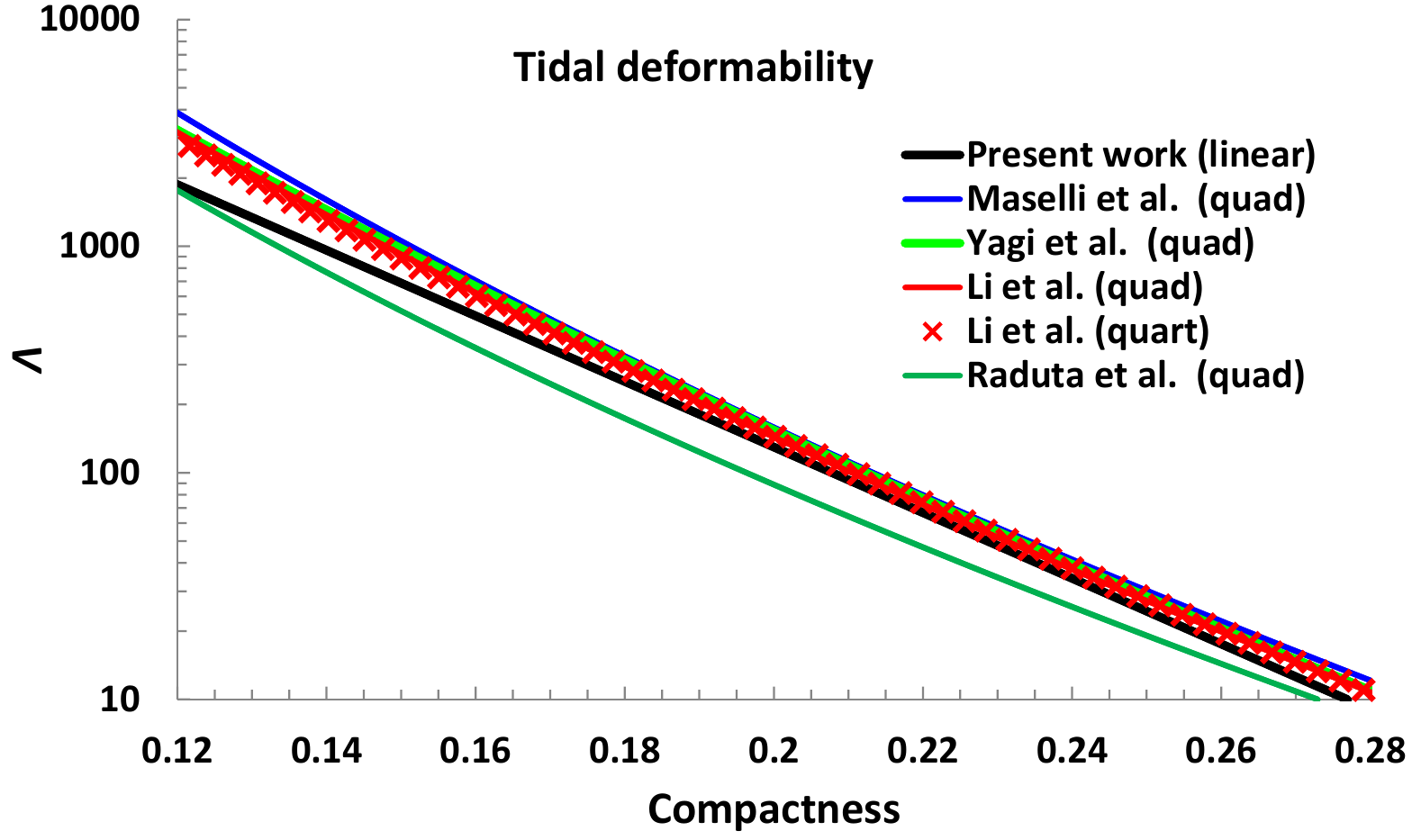}}
	
	\caption{\label{fig:Figure3} (a) and (b) The same as in Figs. \ref{fig:Figure1} (a and b) but for the NS compactness ($\mathpzc{C}$). (c) The tidal deformability (on a logarithmic scale) as a function of $\mathpzc{C}$ for the same EOSs considered in (a). (d) The linear compactness dependence of the tidal deformability of NS ($M\geq\text{M}_{\odot}$) given by Eq.  22, based on both the CDM3Y-Paris and Reid EOSs with ${K}_{0}=230-330$ MeV MeV, in comparison with the equivalent compactness ($\mathpzc{C}$)$-$deformability ($\text{ln}\,\varLambda$) quadratic and quartic relations given by Li \cite{JieSedrakian2023}, Raduta \cite{RadutaOertel2020}, Maselli \cite{MaselliCardoso2013}, and Yagi \cite{YagiYunes2017} \textit{et als}.}
\end{figure*}

In Figures \ref{fig:Fig3a} and \ref{fig:Fig3b} we show the compactness ($\mathpzc{C}$) of NS as a function of the gravitational stellar mass, based on the EOSs considered in Fig. \ref{fig:Figure1}. Figure \ref{fig:Fig3c} exhibits the relation between the NS tidal deformability, on a log scale, and its compactness, based on the same EOSs in \ref{fig:Fig3a}. Figures \ref{fig:Fig3a} and \ref{fig:Fig3b} shows that the stiffer EOS indicates less NS compactness. The indicated compactness of NS of a single solar mass (1 $\text{M}_{\odot}$) decreases from 0.130 (0.129) to 0.119 (0.122) upon increasing the incompressibility of the examined CDM3Y-Paris (Reid) EOS from 200 MeV to 330 MeV. The corresponding compactness of NS (1.4 $\text{M}_{\odot}$) decreases from 0.224 (0.199) to 0.165 (0.167). The increase in compactness with increasing the NS mass goes faster when the indicated maximum NS mass for a given EOS is approached. The effect of the applied EOS on the indicated dependence of $\varLambda$ on $\mathpzc{C}$ appears more clearly in Fig. \ref{fig:Fig3c} in the region of $\varLambda(M\geq\text{M}_{\odot})\geq0.12$, which shows a linear decreasing behavior of log $\varLambda$ with  $\mathpzc{C}$. The slope of this decreasing behavior slightly decreases with increasing the stiffness of the EOS. The origin of this behavior will be explained below. A unified linear fit given by
\begin{equation}
	\label{eqn:fitLC}
	\text{ln}\,(\varLambda) = 11.327 - 32.8\,\mathpzc{C},
\end{equation}
with the best coefficient of determination ${R}^{2}=0.992$, can be used to describe the relation between $\varLambda$ and $\mathpzc{C}$ of NS ($M\geq\text{M}_{\odot}$), as obtained based of the ten CDM3Y-Paris and Reid EOSs with ${K}_{0}=230-330$ MeV. The explicit relation between $\varLambda$ and $\mathpzc{C}$ can be then safely obtained independently of the characteristics of the employed EOS. Typical conclusion and equivalent expressions have been suggested to describe the relation between $\varLambda$ and $\mathpzc{C}$ based on various types of EOSs, but in the form of $\mathpzc{C}$ as a function of ln$(\varLambda)$ up to its quadratic \cite{RadutaOertel2020,YagiYunes2017,MaselliCardoso2013} or quartic \cite{JieSedrakian2023} terms. These studies were performed by applying a post-Newtonian approximation of the two-body metric with an affine description of NS to model the tidal deformability via three EOSs \cite{MaselliCardoso2013}, and with a large collection of Skyrme and other types of EOSs \cite{YagiYunes2017}, in addition to EOSs derived from covariant density functionals and constrained by terrestrial and astrophysical data \cite{JieSedrakian2023} and at fixed values of entropy per baryon and other thermodynamic conditions at non-zero temperature \cite{RadutaOertel2020}. In panel (d) of Fig. \ref{fig:Figure3} we compare the linear expression given by Eq. \ref{eqn:fitLC} and the equivalent quadratic and quartic expressions given in \cite{RadutaOertel2020,YagiYunes2017,MaselliCardoso2013,JieSedrakian2023}, for NS of $M\geq\text{M}_{\odot}$. Figure \ref{fig:Fig3d} shows that the linear expression based on the wide range of both CDM3Y-Paris and CDM3Y-Reid EOSs estimates intermediate values of NS deformability, as a function of its compactness, relative to those estimated by the mentioned studies. Relative to the median values of the six formulae presented in Fig. \ref{fig:Fig3d} at a given $\mathpzc{C}$, the maximum uncertainty of the different formulae decreases from 40\% at $\mathpzc{C}=0.12$ to about 24\% at $\mathpzc{C}=0.27$. The corresponding uncertainty in the present linear formula decreases from 32\% to 12\%. \par
\begin{figure*}[!htbp]
	\centering	
	\subfigure[ \label{fig:Fig4a}]{
		\includegraphics[height=12cm,width=1.0\linewidth]{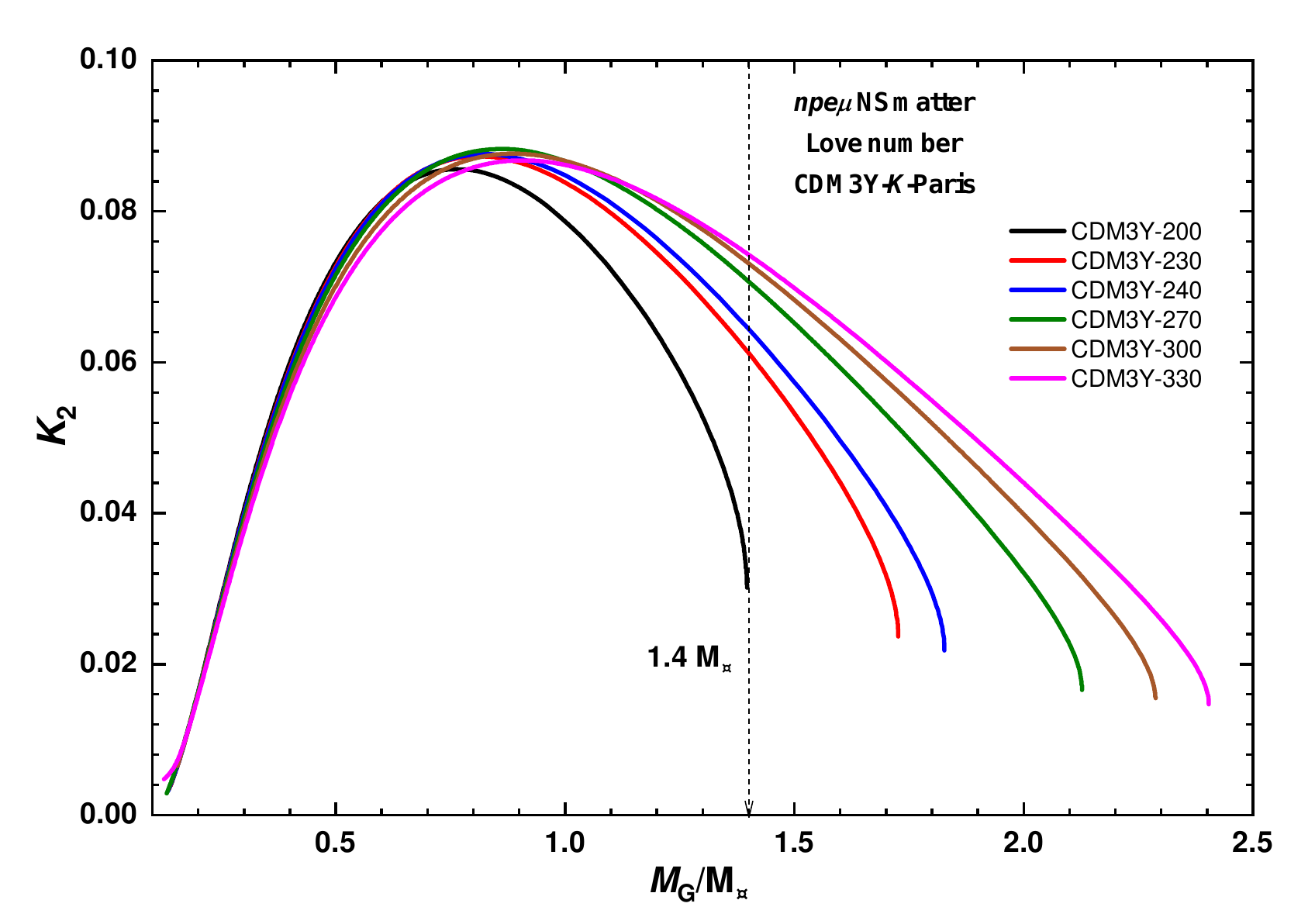}}
	~
	\centering
	\subfigure[ \label{fig:Fig4b}]{
		\includegraphics[height=12cm,width=1.0\linewidth]{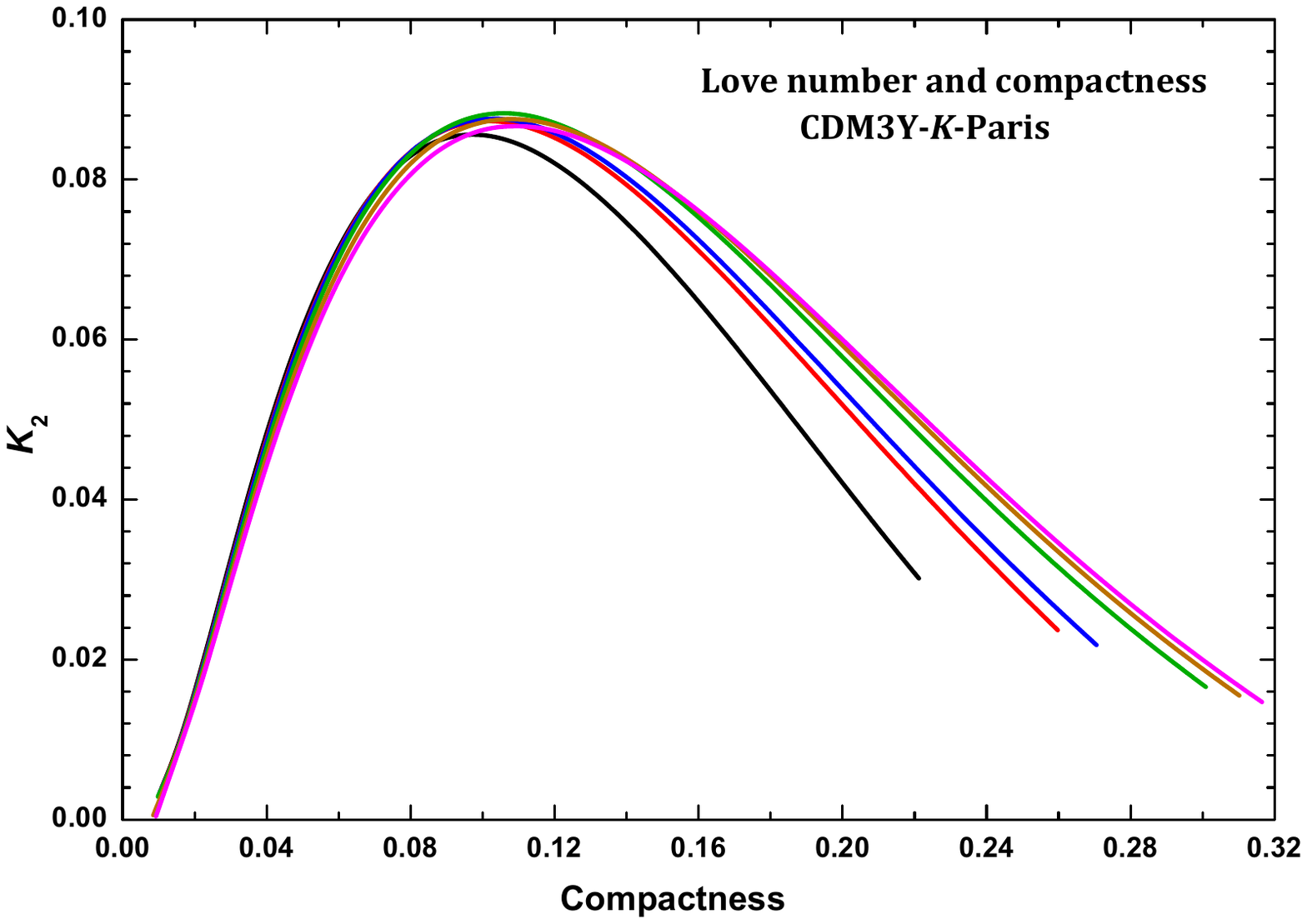}}	
	\caption{\label{fig:Figure4} The same as in Figs. \ref{fig:Fig3a} and \ref{fig:Fig3c} but for the second-order dimensionless tidal Love number (${k}_{2}$).}	
\end{figure*}

The second-order Love number (${k}_{2}$) is displayed in Fig. \ref{fig:Fig4a} as a function of the gravitational stellar mass, based on the same EOSs in Fig. \ref{fig:Fig3a}. The relation between the Love number and compactness of non-rotating spherically symmetric NS as obtained based on the same EOSs is displayed in Fig. \ref{fig:Fig4b}. As seen in Fig. \ref{fig:Fig4a}, both the impact of the stiffness of the EOS on the Love number and its mass dependence vary according to NS mass. While increasing the stiffness of the NS matter slightly decreases the tidal Love number of light NSs of masses less than solar mass, it significantly enhances the love number ${k}_{2}$ for those having masses heavier than the solar mass. For NS of small masses ($M\leq 0.6\, \text{M}_{\odot}$) and small compactness ($\mathpzc{C} \leq 0.08$), ${k}_{2}$ is almost independent of the stiffness of the EOS and its characteristic properties. For such light NS, increasing the crustal width with decreasing the NS mass is expected to take control in decreasing the tidal deformability upon increasing both the compactness and ${k}_{2}$ at a similar rate. For instance, upon decreasing the NS mass from 0.6 to 0.2 $\text{M}_{\odot}$, the estimated compactness (tidal Love number) increases from 0.02 (0.02) to about 0.074 ($0.082\pm0.004$), based on the ten CDM3Y-Paris and Reid EOSs (${K}_{0}=230-330$ MeV). Referring to Eq. \ref{eqn:DIMTDF} and as shown in Fig. \ref{fig:Fig4b}, and the corresponding calculations based on the CDM3Y-Reid EOSs, the similar increasing rates of both ${k}_{2}$ and the most effective $\mathpzc{C}$ strongly decreases the estimated tidal deformability two orders of magnitude within the mentioned range of NS mass, from about $3.4\times{10}^{6}$ (0.6 $\text{M}_{\odot}$) to about $2.5\times{10}^{4}$ (0.2 $\text{M}_{\odot}$), regardless the employed EOS. Based on the same ten EOSs, the tidal love number reaches its maximum value ${k}^{\text{max}}_{2}=0.092\pm0.005$ ($\mathpzc{C}=0.106\pm0.003$) for NS masses of $0.86\pm0.02\,\text{M}_{\odot}$. The indicated tidal Love number turns to decrease with the NS mass heavier than $0.86\pm0.02\,\text{M}_{\odot}$, opposite to continually increasing $\mathpzc{C}$. For the indicated compactness larger than 0.12, which are obtained for the NS masses heavier than $0.97\pm0.04\,\text{M}_{\odot}$, Figure \ref{fig:Fig4b} shows that the calculated Love number exhibits almost ${\mathpzc{C}}^{-1}$ decreasing behavior with the obtained compactness. This turns the tidal deformability given by Eq. \ref{eqn:DIMTDF} to follow ${\mathpzc{C}}^{-6}$ behavior, for NS ($M>\text{M}_{\odot}$). This explains the above described linear behavior of the tidal deformability in the region of $\varLambda(M\geq\text{M}_{\odot})\geq0.12$, when it is plotted on a logarithmic scale, with the corresponding compactness, as seen in Fig. \ref{fig:Fig3c}. The conflict competition between the compactness of NS ($M>\text{M}_{\odot}$) and its corresponding Love number appears as decreasing tidal deformability of such NS with increasing its mass and also with decreasing the stiffness of the employed EOS, as shown in Figs. \ref{fig:Fig1a} and \ref{fig:Fig1b}. The stiffer EOS indicates larger ${k}_{2}$ (less $\mathpzc{C}$) of NS ($M>\text{M}_{\odot}$), leading to enhanced tidal deformability.  
        
\section{\label{sec:level4}SUMMARY AND CONCLUSIONS}
We used recent inferred constraints on mass, radius, and tidal deformability of NS objects, which obtained using restricted schemes from GW and pulsar observations, to inspect the reliability of the semi-microscopic CDM3Y-Paris and Reid EOSs and their compactness to describe highly dense nuclear matter. Two groups of six CDM3Y-${K}_{0}$ EOSs have been used to solve the different structure equations of NS using the TOV equations, from softest CDM3Y-200 EOS characterized with SNM incompressibility of ${K}_{0}=200$ MeV to stiffest one of ${K}_{0}=330$ MeV. We found that the EOSs of 230 MeV$\leq$ ${K}_{0}\leq$ 330 MeV show overall good agreement with the tidal deformability estimated from the GW events, supporting the lower (upper) limits of the indicated constraints of the heavier (lighter) mass component than 1.3 $\text{M}_{\odot}$ of the GW170817 and GW190425 events. \par
For light NS of mass less than $\text{M}_{\odot}$ and relatively large crustal width, the impact of the stiffness of the employed EOS on its tidal deformability is found to be minor. Decreasing the mass of such light NS ($M<\text{M}_{\odot}$) decreases both its indicated compactness and the corresponding tidal Love number, which accordingly indicates enhanced tidal deformability with decreasing its mass. The tidal deformability becomes sensitive to the employed EOS for NS of larger mass than $\text{M}_{\odot}$. The impact of the stiffness of the EOS on the tidal deformability of NS ($M>\text{M}_{\odot}$) is determined by the competition between the NS compactness which decreases upon increasing the NM stiffness and hinders the deformability, and the corresponding less effective Love number which increases with increasing the NM stiffness for the NS of mass larger than the solar mass, and enhances its deformability. The overall behavior appears as an increase in the deformability and its corresponding stellar radius upon increasing the stiffness of the NS matter. The softer EOS indicates a smoother increase in $\varLambda$ with the NS radius. The same situation is obtained for the impact of the NS mass that increases its compactness but hinders the corresponding tidal Love number, which hinder the tidal deformability with increasing the mass of NS ($M>\text{M}_{\odot}$), till it vanishes for the indicated maximum. Unified description is obtained for some correlations between tidal deformability, tidal Love number, and NS compactness independent of the details of the employed EOS. The tidal deformability can then add constraints on the equation of state, but the uncertainty in their experimental estimates still large. Even so, based on a wide range of the CDM3Y EOSs with ${K}_{0}$ between 230 and 330 MeV we obtained narrower ranges of $R$ and $\varLambda$ for NS objects than their experimentally inferred ranges.

\begin{turnpage}
	\begin{table*}[!htbp]
		\centering
		\caption{\label{table1} Radius ($\bm{R}$ \textbf{(km)}) and tidal deformability ($\bm{\varLambda}$) based on the CDM3Y-Paris and Reid EOSs of 230 MeV$\leq$ ${K}_{0}$ (SNM)$\leq$ 330 MeV, for the primary (\textit{$m_1$}) and secondary (\textit{$m_2$}) partners of NS binaries GW170817 and GW190425 of masses listed in column 2, and for NS of indicated maximum mass ${M}_{\text{max}}$ (maximum radius ${R}_{\text{max}}$) based on the considered EOSs, in addition to the NS (1.40 $\text{M}_{\odot}$). The experimentally estimated values of $\bm{R}$ and $\bm{\varLambda}$ are listed in columns 3 and 4. The last three columns show recent calculations of ${M}_{\text{max}}$ and its radius, and $\bm{R}$ and $\bm{\varLambda}$ of NS (1.40 $\text{M}_{\odot}$), using EOSs ($K_0$) based on mean-field calculations \cite{TanKhoa2021,SantosMalik2024, KumarThakur2023, MalikPais2024, LiTian2024, ThakurKumar2022a, ThakurKumar2022b, HuangShen2024}.}    
		\begin{ruledtabular}
			\begin{tabular}{cccccccccc}
				\multicolumn{1}{c}{\multirow{2}{*}{\makecell{\textbf{Object}}}} &
				\multicolumn{3}{c}{\multirow{1}{*}{\textbf{Observed data analysis}}} &
				\multicolumn{1}{c}{\textbf{CDM3Y}} & \multicolumn{2}{c}{\makecell{\textbf{Present calculations}\\$\bm{{K}_{0}=230-330}$ \textbf{MeV}}} & \multicolumn{3}{c}{\multirow{1}{*}{\textbf{Recent calculations}}} \\ \cline{2-10}  	
				&   
				\multicolumn{1}{c}{\multirow{2}{*}{$\bm{M(\text{\textbf{M}}_{\odot})}$}} &
				\multicolumn{1}{c}{\multirow{2}{*}{$\bm{R}$ \textbf{(km)}}} &
				\multicolumn{1}{c}{\multirow{2}{*}{$\bm{\varLambda}$}} & & \multicolumn{1}{c}{\multirow{2}{*}{$\bm{R}$ \textbf{(km)}}} &
				\multicolumn{1}{c}{\multirow{2}{*}{$\bm{\varLambda}$}} & \multicolumn{1}{c}{\multirow{2}{*}{\makecell{$\bm{M(\text{\textbf{M}}_{\odot})}$\\$\bm{{K}_{0}}$ \textbf{(MeV)}}}} & \multicolumn{1}{c}{\multirow{2}{*}{$\bm{R}$ \textbf{(km)}}} & \multicolumn{1}{c}{\multirow{2}{*}{$\bm{\varLambda}$}}     \\
				& & & & & & & & & \\
				\hline
				& & & & & & & & & \\
				\multicolumn{1}{c}{\multirow{2}{*}{GW170817 \textit{m1}}} & \multicolumn{1}{c}{\multirow{2}{*}{\makecell{$1.485\pm0.125$\\\cite{Abbottetal2019,LandryEssick2020}}}} & \multicolumn{1}{c}{\multirow{2}{*}{\makecell{${12.33}^{+1.10}_{-1.45}$\\\cite{LandryEssick2020}}}} & \multicolumn{1}{c}{\multirow{2}{*}{\makecell{$420\pm410$\\\cite{Abbottetal2019,LandryEssick2020}}}} & Paris & $11.72\pm0.85$ & $268\pm204$ & & & \\
				&       &       &       & Reid  & $11.84\pm0.66$ & $305\pm229$ & & &  \\
				& & & & & & & & & \\
				\multicolumn{1}{c}{\multirow{2}{*}{GW170817 \textit{m2}}} & \multicolumn{1}{c}{\multirow{2}{*}{\makecell{$1.250\pm0.110$\\\cite{Abbottetal2019,LandryEssick2020}}}} & \multicolumn{1}{c}{\multirow{2}{*}{\makecell{${12.29}^{+1.20}_{-1.49}$\\\cite{LandryEssick2020}}}} & \multicolumn{1}{c}{\multirow{2}{*}{\makecell{$676\pm605$\\\cite{Abbottetal2019,LandryEssick2020}}}} & Paris & $11.99\pm0.51$ & $732\pm471$ & & & \\
				&       &       &       & Reid  & $12.03\pm0.43$ & $869\pm438$ & & & \\
				& & & & & & & & & \\
				\multicolumn{1}{c}{\multirow{2}{*}{GW190425 \textit{m1}}} & \multicolumn{1}{c}{\multirow{2}{*}{\makecell{$2.045\pm0.445$\\\cite{LandryEssick2020,Abbottetal892L32020}}}} & \multicolumn{1}{c}{\multirow{2}{*}{\makecell{${11.79}^{+1.49}_{-5.21}$ \\\cite{LandryEssick2020}}}} & \multicolumn{1}{c}{\multirow{2}{*}{\makecell{${31}^{+108}_{-31}$\\\cite{LandryEssick2020}}}} & Paris & $11.76\pm0.80$ & $92\pm88$ & & & \\
				&       &       &       & Reid  & $11.82\pm0.71$ & $94\pm93$ & & & \\
				& & & & & & & & & \\
				\multicolumn{1}{c}{\multirow{2}{*}{GW190425 \textit{m2}}} & \multicolumn{1}{c}{\multirow{2}{*}{\makecell{$1.405\pm0.285$\\\cite{LandryEssick2020,Abbottetal892L32020}}}} & \multicolumn{1}{c}{\multirow{2}{*}{\makecell{${12.31}^{+1.11}_{-1.51}$\\\cite{LandryEssick2020}}}} & \multicolumn{1}{c}{\multirow{2}{*}{\makecell{${521}^{+912}_{-461}$\\\cite{LandryEssick2020}}}} & Paris & $11.48\pm1.08$ & $978\pm649$ & & & \\
				&       &       &       & Reid  & $11.78\pm0.75$ & $750\pm703$ & & & \\
				& & & & & & & & & \\
				\multicolumn{1}{c}{\multirow{2}{*}{\makecell{${M}_{\text{max}}$(Paris)\\ $=2.07\pm0.34\,\text{M}_{\odot}$}}}
				&       &       &       & \multicolumn{1}{c}{\multirow{2}{*}{Paris}}   & \multicolumn{1}{c}{\multirow{2}{*}{$10.51\pm0.72$ }} & \multicolumn{1}{c}{\multirow{2}{*}{$11\pm5$}} & & & \\
				& & & & & & & & & \\
				\multicolumn{1}{c}{\multirow{2}{*}{\makecell{${M}_{\text{max}}$(Reid)\\ $=2.13\pm0.25\,\text{M}_{\odot}$}}} &       &       &       & \multicolumn{1}{c}{\multirow{2}{*}{Reid}}  & \multicolumn{1}{c}{\multirow{2}{*}{$10.41\pm0.70$}}  & \multicolumn{1}{c}{\multirow{2}{*}{$4\pm1$}} & \multicolumn{1}{c}{\multirow{2}{*}{\makecell{${M}_{\text{max}}=$\\ 2.222$-$2.573 \cite{SantosMalik2024}}}} & \multicolumn{1}{c}{\multirow{2}{*}{\makecell{$R({M}_{\text{max}})=$\\ 11.11$-$12.41}}} &  \\
				& & & & & & & & & \\
				& & & & & & & & & \\
				\multicolumn{1}{c}{\multirow{2}{*}{\makecell{$M$ (${R}_{\text{max}}$(Paris))\\ $=1.68\,\text{M}_{\odot}$}}}
				&       &       &       & \multicolumn{1}{c}{\multirow{2}{*}{Paris}} & \multicolumn{1}{c}{\multirow{2}{*}{12.56}} & \multicolumn{1}{c}{\multirow{2}{*}{$83\pm53$}} & \multicolumn{1}{c}{\multirow{2}{*}{2.01$-$2.74 \cite{MalikPais2024}}} & \multicolumn{1}{c}{\multirow{2}{*}{10.28$-$13.03}} &  \\
				& & & & & & & & & \\
				\multicolumn{1}{c}{\multirow{2}{*}{\makecell{$M$ (${R}_{\text{max}}$(Reid))\\
							$=1.59\,\text{M}_{\odot}$}}}
				&       &       &       & \multicolumn{1}{c}{\multirow{2}{*}{Reid}}  & \multicolumn{1}{c}{\multirow{2}{*}{12.54}} & \multicolumn{1}{c}{\multirow{2}{*}{$143\pm57$}} & \multicolumn{1}{c}{\multirow{2}{*}{2.01$-$2.19 \cite{ThakurKumar2022a}}} & \multicolumn{1}{c}{\multirow{2}{*}{11.40$-$12.34}} & \\
				& & & & & & & & & \\
				\multicolumn{1}{c}{\multirow{2}{*}{}} & \multicolumn{1}{c}{\multirow{2}{*}{}} & \multicolumn{1}{c}{\multirow{2}{*}{}} & \multicolumn{1}{c}{\multirow{2}{*}{}} & \multicolumn{1}{c}{\multirow{2}{*}{}} & \multicolumn{1}{c}{\multirow{2}{*}{}} & \multicolumn{1}{c}{\multirow{2}{*}{}} & \multicolumn{1}{c}{\multirow{2}{*}{2.34$-$2.77 \cite{KumarThakur2023}}} & \multicolumn{1}{c}{\multirow{2}{*}{11.93$-$12.93}} & \\
				& & & & & & & & & \\
				\multicolumn{1}{c}{\multirow{2}{*}{}} & \multicolumn{1}{c}{\multirow{2}{*}{}} & \multicolumn{1}{c}{\multirow{2}{*}{}} & \multicolumn{1}{c}{\multirow{2}{*}{}} & \multicolumn{1}{c}{\multirow{2}{*}{}} & \multicolumn{1}{c}{\multirow{2}{*}{}} & \multicolumn{1}{c}{\multirow{2}{*}{}} & \multicolumn{1}{c}{\multirow{2}{*}{1.91$-$2.73 \cite{ThakurKumar2022b}}} & \multicolumn{1}{c}{\multirow{2}{*}{11.56 $-$13.22}} & \\
				& & & & & & & & & \\
				\multicolumn{1}{c}{\multirow{2}{*}{}} & \multicolumn{1}{c}{\multirow{2}{*}{}} & \multicolumn{1}{c}{\multirow{2}{*}{}} & \multicolumn{1}{c}{\multirow{2}{*}{}} & \multicolumn{1}{c}{\multirow{2}{*}{}} & \multicolumn{1}{c}{\multirow{2}{*}{}} & \multicolumn{1}{c}{\multirow{2}{*}{}} & \multicolumn{1}{c}{\multirow{2}{*}{1.86$-$2.62 \cite{HuangShen2024}}} & \multicolumn{1}{c}{\multirow{2}{*}{9.16$-$12.59}} & \\
				& & & & & & & & & \\
				
				\multicolumn{1}{c}{\multirow{2}{*}{}} & \multicolumn{1}{c}{\multirow{2}{*}{}} & \multicolumn{1}{c}{\multirow{2}{*}{}} & \multicolumn{1}{c}{\multirow{2}{*}{}} & \multicolumn{1}{c}{\multirow{2}{*}{}} & \multicolumn{1}{c}{\multirow{2}{*}{}} & \multicolumn{1}{c}{\multirow{2}{*}{}} & \multicolumn{1}{c}{\multirow{2}{*}{1.55$-$2.12 \cite{TanKhoa2021}}} & \multicolumn{1}{c}{\multirow{2}{*}{9.30$-$10.40}} & \\
				& & & & & & & & & \\
				\multicolumn{1}{c}{\multirow{2}{*}{$1.4\, \text{M}_{\odot}$}} & \multicolumn{1}{c}{\multirow{2}{*}{}} & \multicolumn{1}{c}{\multirow{2}{*}{\makecell{ ${12.32}^{+1.09}_{-1.47}$ \cite{LandryEssick2020}\\${12.06}^{+1.26}_{-1.43}$ \cite{ThomasRiley2019}}}} & \multicolumn{1}{c}{\multirow{2}{*}{\makecell{${451}^{+241}_{-279}$ \cite{LandryEssick2020}\\${400}^{+259}_{-253}$ \cite{ThomasRiley2019}}}} & Paris & $11.97\pm0.54$ & $311\pm98$ & ${K}_{0}=230\pm40$ & 12.10$-$13.22 & 405$-$662 \cite{SantosMalik2024}   \\
				&       &       &       & Reid  & $12.03\pm0.78$ & $310\pm122$ & 177$-$313 & 11.73$-$13.78 & 291$-$844 \cite{MalikPais2024}   \\
				\multicolumn{1}{c}{\multirow{2}{*}{}} & \multicolumn{1}{c}{\multirow{2}{*}{}} & \multicolumn{1}{c}{\multirow{2}{*}{}} & \multicolumn{1}{c}{\multirow{2}{*}{}} & \multicolumn{1}{c}{\multirow{2}{*}{}} & \multicolumn{1}{c}{\multirow{2}{*}{}} & \multicolumn{1}{c}{\multirow{2}{*}{}} & \multicolumn{1}{c}{\multirow{2}{*}{194$-$279}} & \multicolumn{1}{c}{\multirow{2}{*}{12.02$-$13.40}} & \multicolumn{1}{c}{\multirow{2}{*}{358$-$796 \cite{LiTian2024}}}  \\
				\multicolumn{1}{c}{\multirow{2}{*}{}} & \multicolumn{1}{c}{\multirow{2}{*}{}} & \multicolumn{1}{c}{\multirow{2}{*}{}} & \multicolumn{1}{c}{\multirow{2}{*}{}} & \multicolumn{1}{c}{\multirow{2}{*}{}} & \multicolumn{1}{c}{\multirow{2}{*}{}} & \multicolumn{1}{c}{\multirow{2}{*}{}} & \multicolumn{1}{c}{\multirow{2}{*}{198$-$244}} & \multicolumn{1}{c}{\multirow{2}{*}{12.70 $-$14.07}} & \multicolumn{1}{c}{\multirow{2}{*}{422$-$854 \cite{ThakurKumar2022a}}}  \\
				\multicolumn{1}{c}{\multirow{2}{*}{}} & \multicolumn{1}{c}{\multirow{2}{*}{}} & \multicolumn{1}{c}{\multirow{2}{*}{}} & \multicolumn{1}{c}{\multirow{2}{*}{}} & \multicolumn{1}{c}{\multirow{2}{*}{}} & \multicolumn{1}{c}{\multirow{2}{*}{}} & \multicolumn{1}{c}{\multirow{2}{*}{}} & \multicolumn{1}{c}{\multirow{2}{*}{217$-$270}} & \multicolumn{1}{c}{\multirow{2}{*}{10.60 $-$12.00}} & \multicolumn{1}{c}{\multirow{2}{*}{142$-$407 \cite{TanKhoa2021}}}  \\
				\multicolumn{1}{c}{\multirow{2}{*}{}} & \multicolumn{1}{c}{\multirow{2}{*}{}} & \multicolumn{1}{c}{\multirow{2}{*}{}} & \multicolumn{1}{c}{\multirow{2}{*}{}} & \multicolumn{1}{c}{\multirow{2}{*}{}} & \multicolumn{1}{c}{\multirow{2}{*}{}} & \multicolumn{1}{c}{\multirow{2}{*}{}} & \multicolumn{1}{c}{\multirow{2}{*}{226 $-$272}} & \multicolumn{1}{c}{\multirow{2}{*}{12.96 $-$14.59}} & \multicolumn{1}{c}{\multirow{2}{*}{611 $-$1242 \cite{KumarThakur2023}}}  \\
				\multicolumn{1}{c}{\multirow{2}{*}{}} & \multicolumn{1}{c}{\multirow{2}{*}{}} & \multicolumn{1}{c}{\multirow{2}{*}{}} & \multicolumn{1}{c}{\multirow{2}{*}{}} & \multicolumn{1}{c}{\multirow{2}{*}{}} & \multicolumn{1}{c}{\multirow{2}{*}{}} & \multicolumn{1}{c}{\multirow{2}{*}{}} & \multicolumn{1}{c}{\multirow{2}{*}{223 $-$272}} & \multicolumn{1}{c}{\multirow{2}{*}{12.86 $-$14.65}} & \multicolumn{1}{c}{\multirow{2}{*}{546 $-$1235 \cite{ThakurKumar2022b}}}  \\
				\multicolumn{1}{c}{\multirow{2}{*}{}} & \multicolumn{1}{c}{\multirow{2}{*}{}} & \multicolumn{1}{c}{\multirow{2}{*}{}} & \multicolumn{1}{c}{\multirow{2}{*}{}} & \multicolumn{1}{c}{\multirow{2}{*}{}} & \multicolumn{1}{c}{\multirow{2}{*}{}} & \multicolumn{1}{c}{\multirow{2}{*}{}} & \multicolumn{1}{c}{\multirow{2}{*}{231 $-$292}} & \multicolumn{1}{c}{\multirow{2}{*}{10.74 $-$14.06}} & \multicolumn{1}{c}{\multirow{2}{*}{163 $-$896 \cite{HuangShen2024}}}  \\
				& & & & & & & & & \\
				
			\end{tabular}
		\end{ruledtabular}
	\end{table*}
\end{turnpage} 



\end{document}